\documentclass[twocolumn]{aastex63}
\usepackage{float}
\usepackage{footnote}
\usepackage[toc,page]{appendix}
\usepackage{booktabs}
\usepackage{tabularx}
\usepackage{caption,subcaption}
\usepackage[para,flushleft]{threeparttable}
\usepackage{makecell}
\usepackage{comment}
\usepackage{enumerate}
\usepackage{natbib}
\usepackage{ulem}
\newcommand{\dtoprule}{\specialrule{1pt}{0pt}{0.4pt}%
            \specialrule{0.3pt}{0pt}{\belowrulesep}%
            }
\newcommand{\dbottomrule}{\specialrule{0.3pt}{0pt}{0.4pt}%
            \specialrule{1pt}{0pt}{\belowrulesep}%
            }

\newcommand{\noop}[1]{}
\newcommand{\GG}[1]{}
\received{}
\revised{}
\accepted{}
\submitjournal{ApJ}

\shorttitle{Infrared simulations of star formation regions}
\shortauthors{J\'aquez-Dom\'\i nguez et al.}

\graphicspath{{./}{figures/}}

\begin{document}
\setlength\doublerulesep{0.4pt}
\title{Simulated observations of star formation regions: infrared evolution of globally collapsing clouds}

\correspondingauthor{Jes\'us Miguel J\'aquez Dom\'\i nguez}
\email{j.jaquez@irya.unam.mx}

\author[0000-0002-1912-5394]{Jes\'us M. J\'aquez-Dom\'\i nguez}
\affiliation{Instituto de Radioastronom\'ia y Astrof\'isica, Universidad Nacional Aut\'onoma de M\'exico, Morelia, Michoac\'an 58089, M\'exico.}

\author[0000-0003-1480-4643]{Roberto Galv\'an-Madrid}
\affil{Instituto de Radioastronom\'ia y Astrof\'isica, Universidad Nacional Aut\'onoma de M\'exico, Morelia, Michoac\'an 58089, M\'exico.}

\author[0000-0002-7042-1965]{Jacopo Fritz}
\affiliation{Instituto de Radioastronom\'ia y Astrof\'isica, Universidad Nacional Aut\'onoma de M\'exico, Morelia, Michoac\'an 58089, M\'exico.}

%Other collaborators
\author[0000-0002-2133-9973]{Manuel Zamora-Avilés}
\affiliation{Instituto Nacional de Astrof\'sica,  \'Optica y Electr\'onica, Luis E. Erro 1, 72840 Tonantzintla, Puebla, M\'exico}
\author[0000-0002-4479-4119]{Peter Camps}
\affiliation{Sterrenkundig Observatorium, Universiteit Gent, Krijgslaan 281 S9, 9000 Gent, Belgium}
\author[0000-0002-6971-5755]{Gustavo Bruzual}
\affiliation{Instituto de Radioastronom\'ia y Astrof\'isica, Universidad Nacional Aut\'onoma de M\'exico, Morelia, Michoac\'an 58089, M\'exico.}
\author[0000-0002-3930-2757]{Maarten Baes}
\affiliation{Sterrenkundig Observatorium, Universiteit Gent, Krijgslaan 281 S9, 9000 Gent, Belgium}
\author[0000-0001-9299-5479]{Yuxin Lin}
\affiliation{Max Planck Institute for extraterrestrial Physics, Gießenbachstraße 1,85748 Garching, Bayern, Deutschland}
\author[0000-0002-1424-3543]{Enrique Vázquez-Semadeni}
\affiliation{Instituto de Radioastronom\'ia y Astrof\'isica, Universidad Nacional Aut\'onoma de M\'exico, Morelia, Michoac\'an 58089, M\'exico.}

\begin{abstract}
\noindent 
The direct comparison between hydrodynamical simulations and observations is needed to improve the physics included in the former and test biases in the latter. Post-processing radiative transfer and synthetic observations are now the standard way to do this. We report on the first application of the \texttt{SKIRT} radiative transfer code to simulations of a star-forming cloud. The synthetic observations are then analyzed following traditional observational workflows. We find that in the early stages of the simulation, stellar radiation is inefficient in heating dust to the temperatures observed in Galactic clouds, thus the addition of an interstellar radiation field is necessary. The spectral energy distribution of the cloud settles rather quickly after $\sim3$ Myr of evolution from the onset of star formation, but its morphology continues to evolve for $\sim8$ Myr due to the expansion of \textsc{Hii} regions and the respective creation of cavities, filaments, and ridges. Modeling synthetic \textit{Herschel} fluxes with 1- or 2-component modified black bodies underestimates total dust masses by a factor of $\sim2$. Spatially-resolved fitting recovers up to about $70\%$ of the intrinsic value. This  ``missing mass'' is located in a very cold dust component with temperatures below $10$ K, which does not contribute appreciably to the far-infrared flux. This effect could bias real observations if such dust exists in large amounts. Finally, we tested observational calibrations of the SFR based on infrared fluxes and concluded that they are in agreement when compared to the intrinsic SFR of the simulation averaged over $\sim100$ Myr. 
\end{abstract}


\section{Introduction} 
\label{sec:intro}
Star formation in the Universe is fueled by the gas in molecular clouds (MCs), but many questions remain unanswered. 
Only in recent years the advent of instrumentation that allows to perform multi-scale resolved studies has permitted to 
connect global to local scales \citep[for a review, see][]{Motte18}. An important topic of current research it to link our knowledge of star formation in the Milky Way and external galaxies  
\citep[e.g., the review by][]{Kenn_Evans_2012review}. 
MCs obscure optical starlight and emit themselves from near-infrared (near-IR) to radio wavelengths. 
In the last two decades, infrared (IR) surveys of the Galactic plane using space telescopes such as {\it Spitzer} and {\it Herschel} have unveiled the star formation content of our Galaxy \citep[e.g.,][]{Churchwell09,Molinari10}. Complementary ground-based surveys in the (sub)millimeter have been key to map the bulk of the mass in MCs as traced by their cold ($T \sim 10$ to 30 K) dust and molecular gas \citep[e.g.,][]{Schuller09,Aguirre11}.     

Important observational constraints now exist on the physics of star formation. Among the most important ones is that Galactic regions have been inferred to be inefficient in converting gas into stars, except possibly in their densest parts. 
The inferred -- current -- star formation efficiencies range from about zero to a few percent in nearby clouds \citep[e.g,][]{Forbrich09,Evans09} to $> 10\%$ for the most active clouds in the Milky Way \citep[e.g.,][]{galvan2013muscle,Louvet14,Ginsburg2016}.
Several theoretical scenarios for MC evolution and star formation have been able to reproduce a variety observational restrictions \citep[see, e.g.,][]{Federrath_2012,Krumholz2012,vazquez2019global,smith2020cloud,Hennebelle2022}. Among them, simulations of MCs that are under global collapse are particularly appealing for the regime of massive star formation in clustered environments  \citep[e.g.,][]{vazquez2007molecular,Heitsch_2008,Colin_2013,Ibanes_2016,ZA+19} since massive star-forming clumps appear to be dominated by self-gravity \citep[e.g.,][]{LiuBaobab2012,LinYuxin2016,LiuBaobab2017}. 
%Ideally, these simulations should have the effect of ionizing feedback from young massive stars, which is the main ingredient thought to stop the otherwise runaway star formation and set the final star formation efficiency \citep[e.g.,][]{Matzner_2002,Peters2010,Dale2012,Haid_2019}. 
Ideally, such simulations should include the effect of ionizing feedback from young massive stars, which is one of the main ingredients responsible for stopping the otherwise runaway star formation and set the final star formation efficiency \citep[e.g.,][]{Matzner_2002,Peters2010,Dale2012,Haid_2019}.
Other factors that are important to determine the final star formation efficiency of a MC are initial conditions, such as  cloud-scale magnetic fields and the amount of turbulence set by large-scale supernovae \citep[e.g.,][]{MK2004,Commercon2011,Peters2011,Federrath_2012}, as well as other types of feedback such as stellar winds, bipolar outflows, and radiation pressure \citep[e.g.,][]{Dale2014,Geen2021,Olivier2021,Rosen2022}. 

However, theoretical models need further testing. An important limitation for a correct comparison between hydrodynamical simulations and observations is the barrier imposed by radiative transfer and instrument response. The post-processing calculation of the propagation of radiation across the MC and to the observer for the production of synthetic observations is becoming the standard technique to homogeneously compare simulations and observations \citep[e.g,][]{Arthur2011,koepferl17a,Betti_2021,Izquierdo_2021RTCloudFactory}. A recent review on this topic is in \citet{Haworth218_review}. 
The comparison of synthetic observations of dust emission from the near to the far-IR is particularly useful because dust continuum emission is the most widely-used tracer of the interstellar medium \citep[e.g.,][]{koepferl17b,Reissl2020,LiuJunhao2021}. Such synthetic observations can be used to test how accurate are hydrodynamical simulations in recreating the physical properties of star formation regions.
Conversely, synthetic observations can be used to test the observational techniques commonly used to calculate the properties of MCs. The comparison between the actual properties of the simulated clouds and the derived ones  can be used to quantify the limitations and biases affecting these techniques.

In this work, we create and analyze radiative-transfer synthetic observations of the simulations presented by \citet{ZA+19}. These models belong to what is known as the Global Hierarchical Collapse (GHC) scenario of MC formation and evolution \citep[][]{Vazquez_2009,vazquez2019global}. The simulation forms a young star cluster with massive star formation and ionization feedback. In Section \ref{sec:simulations} we describe the hydrodynamical simulation. In Section \ref{sec:RTmodels} we explain the methodology for the dust continuum radiative transfer. In Section \ref{sec:synthetic_observations} we outline the production of synthetic observations. In Section \ref{sec:results} we present our results, including a comparison to real observations. Finally, we discuss our results in  Section \ref{sec:discussion} and give our conclusions in  Section  \ref{sec:conclusions}.

%%%%%%%%%%%%% hydrodynamic_simulation
\section{Hydrodynamical simulation}
\label{sec:simulations}
The snapshots used in this work are taken from the radiation magnetohydrodynamical simulation presented in \cite{ZA+19}, which follows the self-consistent formation and collapse of a cold molecular cloud from colliding flows in the warm neutral medium (WNM). We refer the reader to \cite{ZA+18} and \cite{ZA+19} for a complete description of the numerical methods in the hydrodynamical simulations. Here we briefly describe their setup.

The hydrodynamical simulation was performed in {\tt FLASH} \citep[V2.5;][]{fryxell2000}. It includes magnetic fields, self-gravity, sink particles, and feedback by ionizing photons. 
The ionization feedback implemented in FLASH uses the ``hybrid characteristics'' approach adapted from  \citet{Peters2010}, which builds on the work by \citet{Rijkhorst2006}. First, the hydrogen column density is calculated along rays from the sink particles, and the rate equation for the ionization fraction of the cells is solved along these rays. Heating and cooling are calculated separately for ionizing and non-ionizing radiation. We refer the reader to \citet{ZA+19} for a more detailed description of these implementations.

The size of the numerical box is of $256 ~ \mathrm{pc}$ in the $x$ axis, and $128 ~ \mathrm{pc}$ in the $y$ and $z$ axes. The box initially contains warm neutral gas with uniform density and temperature of $n = 2 ~ \mathrm{cm}^{-3}$ and $T = 1450 ~ \mathrm{K}$. The initial composition is assumed to be purely atomic gas with molecular weight $\mu = 1.27$. Thus, the total mass within the box is $M_\mathrm{tot} \approx 2.6 \times 10^5 ~ M_\odot$. The grid refinement criterion employed is one of ``constant mass'', instead of the standard Jeans criterion. This means that, rather than maintaining a constant number of cells per Jeans length (which implies $\Delta x \propto \rho_\mathrm{thr}^{1/2}$) we impose that the cell's mass is the same at each new refinement step (implying $\Delta x \propto \rho_\mathrm{thr}^{1/3}$), and the maximum physical resolution reached is $\Delta x = 0.03 ~ \mathrm{pc}$. 

\begin{figure}[htb!]
    \includegraphics[width =0.47\textwidth, height =0.35\textheight]{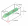}
    \caption{Sketch illustrating the initial conditions used in the simulations presented by \cite{ZA+19}, which consist of two cylindrical streams colliding at the center of the numerical box.} 
    \label{fig:Manuel_setup}
\end{figure}

The molecular cloud is formed by compression of two cylindrical flows from the WNM, as schematically depicted in \autoref{fig:Manuel_setup}. The flows have radius $R_{ \mathrm{flow}} = 32 ~ \mathrm{pc}$ and length $L_{\mathrm {flow}} = 112 ~ \mathrm{pc}$, and move in opposite directions along the $x$ axis with a supersonic velocity $v_{\rm flow} = 7.5 ~ \mathrm{km ~ s^{-1}}$. A background subsonic velocity field with a $k^{-2}$ spectrum is imposed in the box with Mach number $\mathcal{M}_\mathrm{rms} \approx 0.7$. Finally, the box is permeated by a magnetic field aligned along the $x$ direction, with an initial uniform intensity of $3 ~ \mathrm{\mu G}$, consistent with observations \citep{Beck01}.

%check V1 JF

Sink particles are formed in the simulation once a cell reaches the maximum level of refinement and the gas density within it is larger than a threshold value of $\rho = 8.9 \times 10^{-18} ~ \mathrm{g ~ cm^{-3}}$. 
The gaseous cell retains the excess mass above the density threshold, i.e., the initial sink particle mass is  
$M_\mathrm{sink} = (\rho - \rho_\mathrm{thr}) \Delta x^3$. Then, the sink will continue to accrete mass from its surroundings within an accretion radius $\sim 2.5 \Delta x$. Once the sink particles are massive enough, they start to inject ionizing photons into their surroundings, creating \textsc{Hii} regions and sweeping out their surrounding gas.  

The simulation snapshots analyzed in the radiative transfer post-processing presented here, are selected after the cloud undergoes a transition to the cold atomic phase, begins to contract gravitationally, and attains the temperature and density conditions typical of cold molecular clouds. The snapshots are selected to sample the following values of the cloud's instantaneous star formation efficiency (SFE), defined as the ratio of the mass in sinks over the total mass in sinks and gas. 
These correspond to values of the SFE from relatively quiescent clouds \citep[e.g.,][]{Battersby14,Wang2014,lin2017cloud} to active massive star formation regions \citep[e.g.,][]{LinYuxin2016,Motte2022,Traficante2023}. The selected snapshots correspond to times $t = [0.8, 1.6, 3.3, 5.7, 6.3, 7.6]$ Myr after the first star is born, with the highest value approximately corresponding to the moment when feedback by supernovae  might become relevant. This type of feedback is not included in the hydrodynamical  simulation.  
%%%%%%%%%%%%%%%%%%%%%%%% Raditaive transfer 

\section{Radiative transfer}
\label{sec:RTmodels}
The post-processing radiative transfer (RT) was carried out with the Monte Carlo code \texttt{SKIRT}\footnote{\url{https://www.skirt.ugent.be}}  version 9  \citep{camps2015skirt,camps2020SKIRT}. 
This code includes the full treatment of absorption and multiple anisotropic scattering by dust and calculates the thermal emission self-consistently, also including the stochastic heating of the smallest dust grains. The code handles arbitrary three-dimensional (3D) geometry for both radiation sources and dust distributions, including grid-based representations generated by hydrodynamical simulations.
In recent years,  \texttt{SKIRT} has been used mostly to generate synthetic observations from hydrodynamically simulated galaxies \citep[e.g.,][]{Camps_2016SKIRT,Elagali_2018SKIRT,Williams_2019SKIRT,Parsotan_2021SKIRT,Camps_2022SKIRT,Nanni_2022SKIRT}. To our knowledge, this work is the first effort to expand the use of  \texttt{SKIRT} to  simulations of Galactic molecular clouds.

In \texttt{SKIRT}, a radiative transfer model is defined by the following components:
\textit {a)} the source system defined by the SED and spatial distribution of each of the primary sources of radiation;  
\textit {b)} the dust system, in which the physical properties, composition, and spatial distribution of the dust are described; and finally, 
\textit {c)} the synthetic instruments, which determine the wavelengths, position, and properties of the synthetic detectors. 
One of the improvements of \texttt{SKIRT} in version 9 is the possibility of having different  wavelength grids for each component of the radiative transfer simulation. This allows to optimize the use of photon  packages at the specific wavelengths where they are more important.

\subsection{Source system}
\label{subsec:sinks}
The hydrodynamical simulation does not resolve the formation of individual stars. Therefore, the sink particles have to be treated as stellar aggregates. To obtain the SED for the sources of stellar radiation we use the revised version of the \citet{bruzual2003stellar} stellar population synthesis models introduced in \citet[][hereafter C\&B models]{plat2019}. The output SED of the stellar aggregate representing each sink particle depends on its total mass, age, metallicity, and the choice of a stellar initial mass function (IMF). The total mass, age, and position of each sink particle are known directly from the hydrodynamical simulation, while the metallicity is assumed to be solar. The C\&B models make use of the PARSEC evolutionary tracks \citep{Bressan2012_PARSEC}. Finally, we adopt a \citet{kroupa2001variation} IMF with lower- and upper-mass limits of 0.1 and $100 ~M_\odot$, respectively. 

An important point to keep in mind in the construction of the stellar aggregates is that the masses of the sink particles (up to a few $\times 10^2~M_\odot$) are not enough to fully sample the IMF. Thus, they cannot be considered simple stellar populations. For each sink particle, we need to distribute their total mass into individual stars using the IMF as a probability distribution function  \citep[see, e.g.,][]{bruzual2010}. Due to the stochastic nature of populating the sinks in this way, any time that we create a realization of the stellar population for a sink, the SED will be different.  
To deal with these biases, for each sink particle we considered $\sim 200$ realizations of a stochastically-sampled IMF. We then produced the SED of each realization and calculated their bolometric luminosity. Finally, we used the spectrum with a luminosity closest to the median as the input SED for each sink particle at each snapshot.

\begin{figure}
    \centering
    \includegraphics[height =0.33 \textheight]{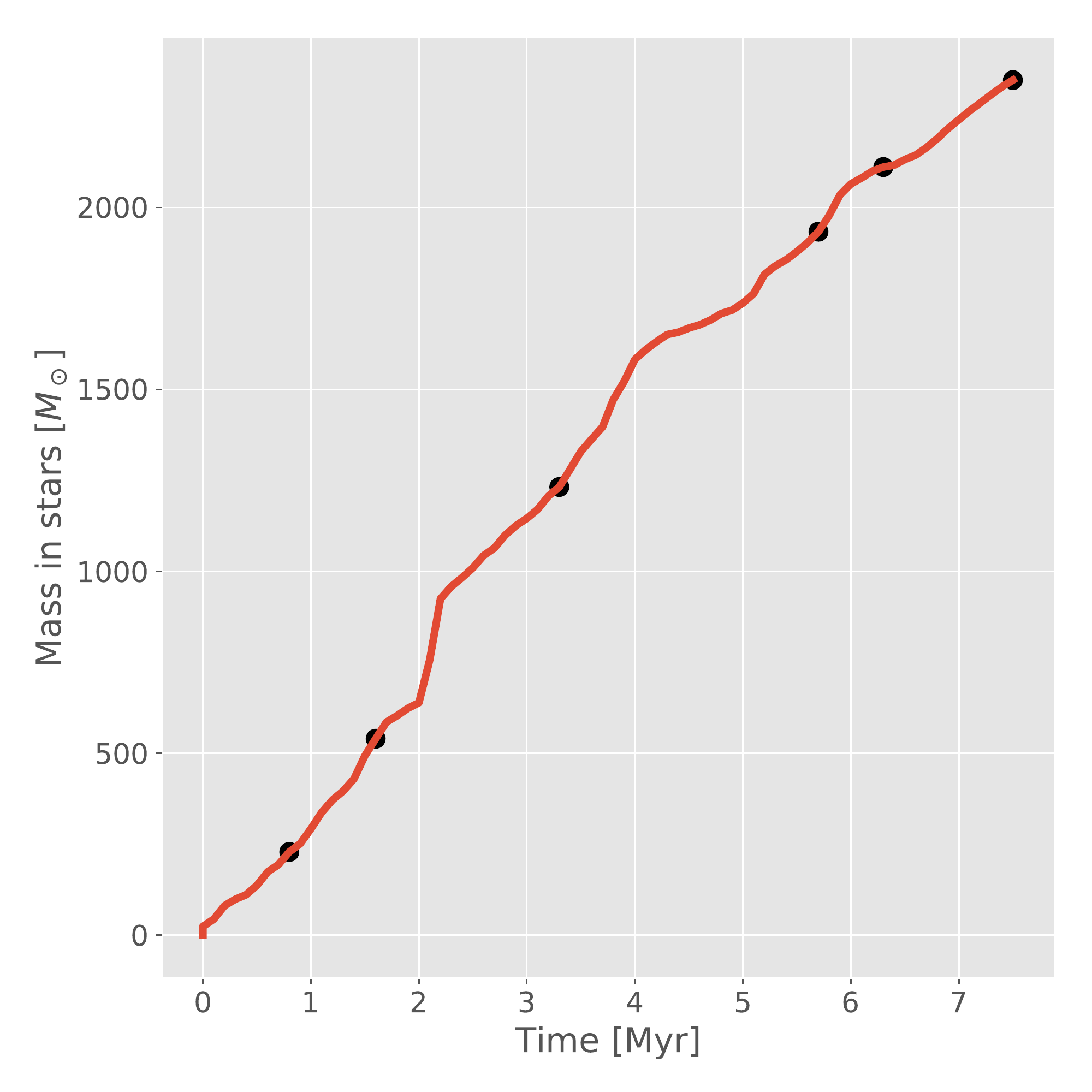}
    \caption{Evolution of the total mass in stars in the hydrodynamical  simulation. The black dots show the age of the snapshots for which we have calculated the radiative transfer and created synthetic observations.}
    \label{fig:Mass_evolution}
\end{figure}

Figure \ref{fig:Mass_evolution} shows the cumulative mass locked in sinks in the simulation as a function of time. Our radiative transfer simulations cover snapshots from a quiescent stage with little star formation up to an evolved massive stage with a total stellar mass of a few $\times 10^3~M_\odot$, which is comparable to the total mass in young stars in a molecular cloud such as Orion \citep{DaRio14,Stutz18}.

Once we have characterized the SEDs of the sinks and their bolometric luminosities, we proceed to give them as input to \texttt{SKIRT}. The code distributes the total energy into photon packages with wavelengths between 0.01 to $2000 ~ \mathrm{\mu m}$ following a logarithmic spacing. These photons are then isotropically launched from the position of the sinks to the medium. The wavelength limits are chosen in such a way to include UV and optical photons that are the most important for dust heating, as well as IR emission from the primary sources of radiation. 

Finally, we define the grid for the \textit{radiation field}, which defines the wavelength points used to record the energy deposited by photon packets that interact with the dust in each cell. For this, we defined a grid with 50 bins distributed logarithmically in the ranges from $0.01$ to $160~\mathrm{\mu m}$. This allowed to optimize the use of computer memory, since the contribution to the dust heating due to photons with a wavelength greater than $160~\mathrm{\mu m}$ is negligible. We refer the reader to section 5.4 of \citet{camps2020SKIRT} for a more detailed description and justification of the wavelength grids.

\subsection{Dust Model}
\texttt{SKIRT} allows a flexible modelling of the dust composition. We chose the multi-component dust mix of \cite{draine2007infrared}, which is appropriate for Milky Way dust. 
This dust mix is composed of silicates, -- small -- graphite, neutral and ionized polycyclic aromatic hydrocarbons (PAHs). The grain size range goes from $3.1 \times 10^{-4} ~ \mathrm{\mu m}$ to $2 ~ \mathrm{\mu m}$. The first two types follow a power-law size distribution, and the last one a log-normal. For each grain component, we use 15 size bins to discretize the calculations of the dust mix. 

The gas cells from the hydrodynamical  simulation are imported into \texttt{SKIRT}, which then converts the gas grid into a dust grid. We define the amount of dust in each cell $M_\mathrm{dust}$ as the amount of non-ionized gas multiplied by the dust-to-gas ratio, which we take as 0.01 \citep{Draine2011Book}: 
\begin{equation}
    M_\mathrm{dust} = 0.01 M_\mathrm{gas} (1 - F_\mathrm{ion}),
    \label{cell_dust_mass}
\end{equation}
\noindent
where $F_\mathrm{ion}$ is the ionization fraction of the cell.

\subsection{Synthetic instruments} 
\label{sec:wavelength_grid}
The first instrument that we configured emulates an ideal spectrometer. The wavelength grid for this instrument is defined in a nested  logarithmic way. The basic grid goes from $0.01$ to $3000~\mathrm{\mu m}$ and is divided into $250$ bins. A high resolution sub-grid with $30$ wavelength bins was also defined spanning the range from $1$ to $35~\mathrm{\mu m}$, in order to better sample PAH emission and the 9.7 $\mu$m silicate absorption feature.

The second instrument that we used simulates a photometric camera that captures the surface brightness resulting from the Monte Carlo radiative transfer. We configured this camera with $4000 \times 4000$ pixels. The distance to the instruments was set to $5~\mathrm{kpc}$, which is an appropriate value for typical massive star formation regions in the Galaxy \citep{Reid14}. This corresponds to an angular pixel size of $1.32\arcsec$ (0.032 pc), or about the maximum physical resolution of the hydrodynamical simulation. The total field of view of this camera is $88 \times 88 ~ \mathrm{arcmin}^2$ ($128 ~ \mathrm{pc}$ per side), which ensures the full coverage of the central part of the simulation (see Figure \ref{fig:Manuel_setup}).

Although the radiative transfer models are designed to obtain results in a wide range of the spectrum (UV-radio), in this work we only analyze observations in the infrared,  mainly to simulate photometric instruments of the \textit{Spitzer} and \textit{Herschel} observatories.

\subsection{ISRF model} \label{sec:ISRF}
The earlier snapshots in which most of the stars are not yet born, or are still deeply embedded, should have in principle properties similar to those of the infrared dark clouds (IRDCs) observed in the Galaxy \citep{Rathborne2006}. These IRDCs have gas and dust temperatures as low as $\approx 10$ K \citep{Pillai11,Battersby14,Wang2014,lin2017cloud}.  However, the hydrodynamical simulation does not include the sources that produce early dust heating in the molecular cloud. To achieve realistic initial temperatures, we include an additional source of photons simulating a kind of interstellar radiation field (ISRF). This ISRF is included in the form of a shell of $64~\mathrm{pc}$ radius that radiates inward only. The SED of the ISRF is constructed using the parametrization of \cite{hocuk2017parameterizing}. This consists of the sum of six modified black bodies \citep{zucconi2001dust} plus an UV contribution \citep{draine1978photoelectric}. Each of the blackbody parts can be characterized as:
\begin{equation} \label{eq:J-no-UV}
    J_\nu^\mathrm{no-UV} = \frac{2h \nu^3}{c^2} \sum_i \frac{W_i}{\exp{(h\nu / K_B T_i)} -1},
\end{equation} 
\noindent 
where the values for $W_i$ and $T_i$ are given in Table \ref{tab:wi_ti}. The UV part of the spectrum is adopted from \cite{draine1978photoelectric}, rewritten in the following form to match units:  %\ref{tab:temp_weighed_mass}
\begin{equation}  \label{eq:J-UV}
    J_\nu^\mathrm{UV} = 4280(h \nu)^2 - 3.47 \times 10^{14}(h \nu)^3 + 6.96 \times 10^{24}(h \nu)^4.
\end{equation}
The combined radiation field is given by: 
\begin{equation} \label{eq:J-ISRF}
J_\nu^\mathrm{ISRF} = J_\nu^\mathrm{no-UV} + J_\nu^\mathrm{UV}.  
\end{equation} 

In Eqs. \ref{eq:J-no-UV}, \ref{eq:J-UV}, and \ref{eq:J-ISRF} the mean intensities are in $\rm{erg \ s^{-1} \ cm^{-2} \ Hz^{-1}  \ sr^{-1}}$, and all other quantities are also in cgs units.

\begin{table}
\centering
\caption{Parameters for the six modified black bodies from the non-UV ISRF.}
\begin{tabular}{c  c  c }
\dtoprule
$\lambda [\mu m]$ & $W_i$ & $T_i ~$K \\ \hline
0.4  & $1 \times 10^{-14}$ & 7500 \\
0.75 & $1 \times 10^{-13}$ & 4000 \\
1    & $4 \times 10^{-13}$ & 3000 \\
10   & $3.4 \times 10^{-9}$ & 250 \\
140  & $2 \times 10^{-4}$ & 23.3 \\
1060 & 1  & 2.728 \\ \dtoprule
\end{tabular}
\label{tab:wi_ti}
\end{table}

The impact of not including the ISRF as an extra source of radiation is more evident in snapshots previous to the formation of massive sinks. In these early time steps, \textsc{Hii} regions have not started to sweep their surrounding dense gas, and the UV-optical photons cannot escape to distances beyond a fraction of a pc. Without the ISRF which permeates the cloud from the outside, in these early snapshots the stellar radiation field is reprocessed only by dust in the immediate surroundings of the -- fewer -- young stars. Therefore, the photons that exit from these compact dust cores are predominantly at IR wavelengths, thus are inefficient to heat the dust across the entire cloud. The result of this is that the bulk of the  cloud mass is kept at temperatures of a few Kelvin, which is unrealistically low.  

As we described in Section \ref{subsec:sinks}, the sources in \texttt{SKIRT} are defined by the shape of the SED and its normalization, which in this work we are using the bolometric luminosity. The form of the ISRF has already been defined using the Equation \ref{eq:J-ISRF}.
Then, taking advantage of the fact that we can define the bolometric luminosity by hand, we perform different simulations where we vary the value of the ISRF luminosity in order to simulate that our cloud is surrounded by different environments and thus see how the temperature of our cloud changes.
We generate three different models for the earlier 0.8 Myr snapshot. To quantify the difference between models we use the mass-weighted average temperature, which is calculated by multiplying the mass and the temperature over the grid. The first model, which we call STARS, uses only the sinks as photon sources. The total sink luminosity input for this model is  $L_\mathrm{stars} = 1.47 \times 10^4~L_\odot$. The resulting mass-weighted average temperature across the cloud is only 3.73 K. For the next model, ISRF, we only use the ISRF as a source of radiation. The ISRF luminosity is $L_\mathrm{ISRF}=10^4~L_\odot$. The resulting mass-weighted temperature is 6.21 K. Finally, the model where we include both sources of radiation, ISRF+STARS, has resulting mass-weighted temperature is 8.69 K, which is more consistent with the $\sim 10$ K expected from observations. 

The previous test illustrates our point that the ISRF is needed, and indeed it dominates dust heating across the cloud during earlier stages of the simulation prior to significant stellar feedback. In Table \ref{tab:temp_weighed_mass} we summarize the results of these tests. For the model where we include the ISRF and stars, we test decreasing the total ISRF luminosity to $\sim 10^3~L_\odot$ and found that the mass-weighted average cloud temperature is reduced to $4.55$ K, while if we increase to values much larger than $10^4~L_\odot$ does not result in a significant increase in dust temperature. 
Therefore, we fix $L_{\mathrm{ISRF}} = 10^4~L_\odot$. 

\begin{table}[t]
\caption{Luminosity input of the primary radiation sources in the radiative transfer models and the resulting mass-weighted dust temperature for the cloud in the earliest snapshot at 0.8 Myr.} 
    \centering
    \begin{tabular}{c c c}
        \hline
        \hline
        Model & Luminosity  & Dust temperature\\ 
        0.8 Myr & [$L_\odot$] & [K] \\
        \hline
        STARS  & $1.47 \times 10^4$ & 3.73 \\
        ISRF & $ 10^4$ & 6.21 \\
        ISRF+ STARS & $2.47 \times 10^4$ & 8.69 \\ \hline     
    \end{tabular}
    \label{tab:temp_weighed_mass}
\end{table}
%%%%%%%%%%%%%%%%%%%%%%%%% fin Radiative transfer
%%%%%%%%%%%%%%%%%%%%%%%%% begin Synthetic observations
\begin{figure*}
    \captionsetup[subfigure]{labelformat=empty}
   \centering
   \begin{subfigure}{\textwidth}\includegraphics[height=0.4\textheight, width =\textwidth]{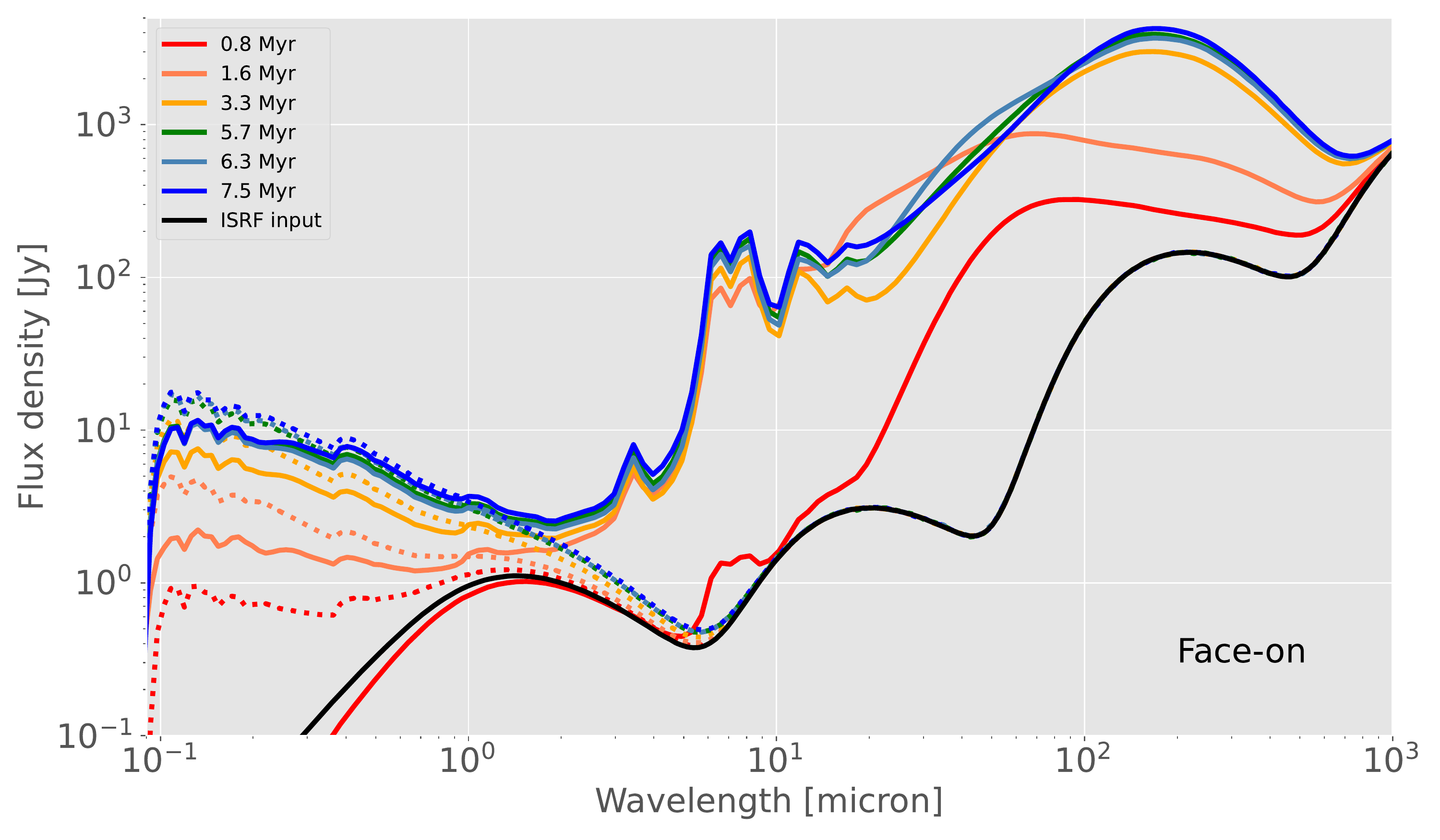}
   \caption{}
   \end{subfigure}
   \begin{subfigure}{\textwidth}
    \includegraphics[height=0.4\textheight, width =\textwidth]{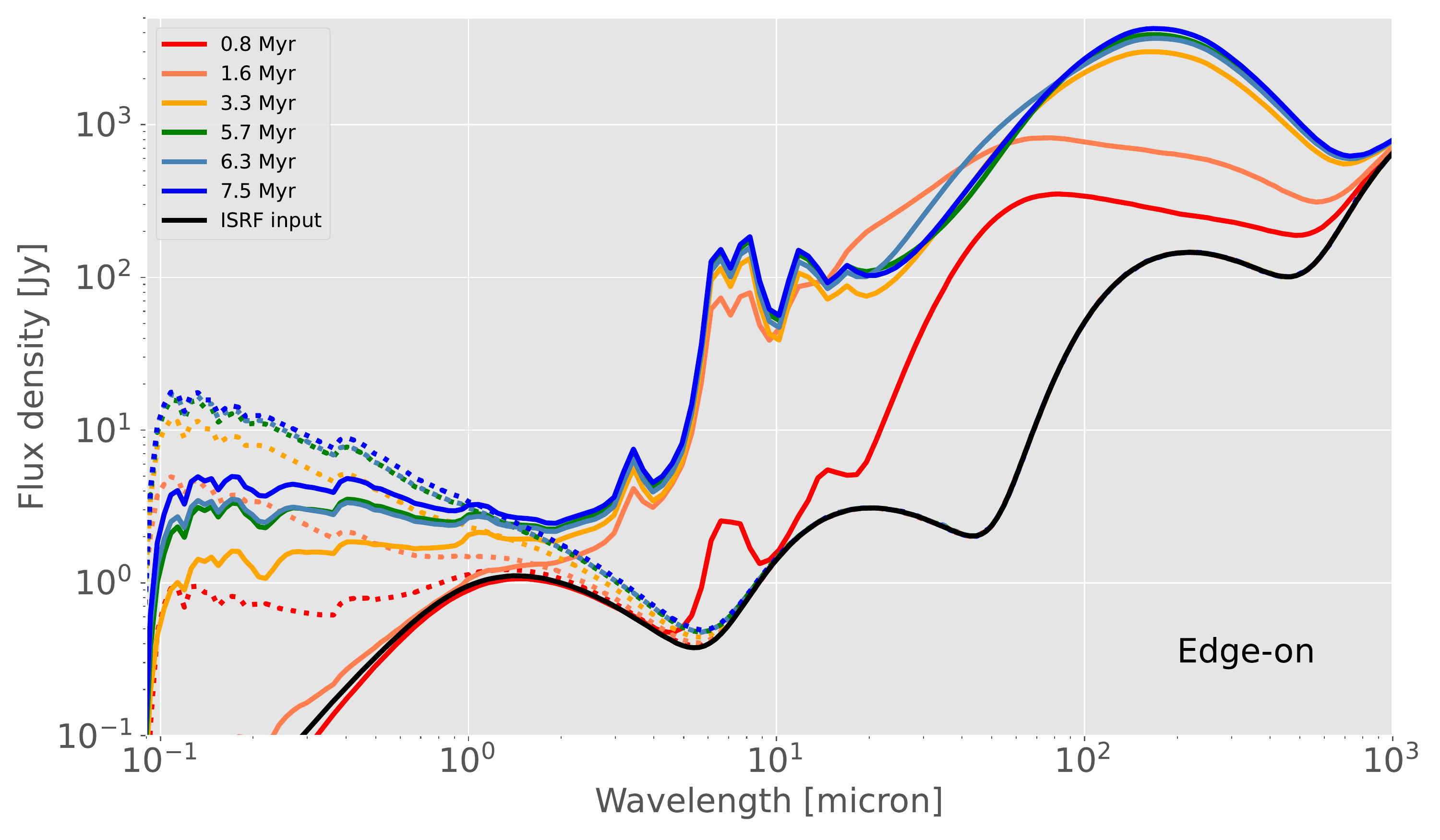}
    \caption{}
    \end{subfigure}
    \caption{SEDs resulting from the ideal synthetic observations. The upper panel shows the SED observed for the face-on line of sight. The lower panel shows the edge-on view. The colored full lines show the resulting cloud SED for the different snapshots. The dotted lines represent the respective total SED (stars + ISRF) of the input sources. The solid black line highlights the input SED of the ISRF source alone.}
    \label{fig:total_SED}
\end{figure*}

\section{Synthetic observations} 
\label{sec:synthetic_observations}
\subsection{Synthetic SED}
\label{subsec:synthetic_raw_SED}
In this section, we describe the raw SEDs resulting from the radiative transfer simulations. For this, we analyze the output SED from the instrument that simulates an ideal spectrometer, which captures the photons that leave the region in the whole spectral range. This instrument has an infinite aperture and is configured with the wavelength grid defined in Section \ref{sec:wavelength_grid}. 
Figure \ref{fig:total_SED} shows the resulting ideal SED for the face-on and edge-on views\footnote{The face-on and edge-on views are defined along the x and y axes of Figure \ref{fig:Manuel_setup}, respectively.} of the cloud for the different timesteps, as well as the input-source SEDs.

The first interesting feature that stands out in both lines of sight, is seen in the SED of the 0.8 Myr snapshot (red line), which shows very low UV-optical emission (at 0.1 to $1 ~ \mathrm{\mu m}$). Instead of the typical stellar emission (dotted line), we see a ``blackbody shape" that peaks between 1 to 2 $\mu m$ in near-IR H band. Comparing the cloud SED with the input ISRF we can see that almost all of the cloud emission at shorter wavelengths is due to the latter. 
Emission that is reprocessed by the dust in the cloud, both from the ISRF and the few young stars, shines as a second peak in the cloud SED at $\sim 70 ~\mu$m (see Fig. \ref{fig:total_SED}).
The cloud emission toward longer wavelengths (100 to $1000~\mu m$) comes from dust far from the input sources that remains at lower temperatures, as well as from photons from this part of the ISRF spectrum that do not interact with the ISM. Strictly speaking, the latter contribution to the observed SED is not cloud emission. In our observational treatment of the synthetic images we eliminate this artefact by means of background subtraction (see Section \ref{subsec:synthetic_broadband_images}). 
Finally, characteristic emission from PAHs is visible at around $6.2~\mu$m, as well as the silicate feature at $9.7 ~ \mathrm{\mu m}$.

%% second timestep
In the second time step (1.6 Myr) we see that stellar emission at $< 1~ \mathrm{\mu m}$ starts to contribute to the SED. This is more evident in the face-on view, for which the dust surface density in the line of sight is smaller with respect to edge-on. In both lines of sight, PAH emission and silicate bands at $\sim 3.6$, $6.2$, and $9.7 ~ \mathrm{\mu m}$ are all now well-defined. An interesting feature of this SED is the shape in the mid- to far-IR wavelengths range ($\sim 20$ to $500 ~\mu$m), where emission is nearly flat in the log-log representation. This indicates that the dust in the cloud has a wide range of temperatures from $\sim 10$ to 100 K similarly contributing to the observed flux. 

For the remaining snapshots at $3.3$, $5.7$, $6.3$, and $7.5$ Myr, the stellar emission becomes more significant as a consequence of stellar-mass growth (see Fig. \ref{fig:Mass_evolution}). The effect of  \textsc{Hii} region feedback, which sweeps up material exposing the stellar sources, is also a contributing factor. As the cloud evolves, more sinks are formed and the protostellar cores are dispersed, allowing more stellar UV and optical photons to reach the instrument. The PAH features remain prominent and the global peak of the SED remains in the far-IR around at $\sim 160~ \mu$m, dominated by dust at $\sim 20$ K. Overall, the cloud SED converges to an almost constant shape after $\sim 3$ Myr of evolution from the onset of star formation,  as marked by the first appearance of sink particles. 
This timescale is appropriately bounded in the lower end by the $\sim 1$ Myr of the added lifetimes of massive young stellar objects and ultracompact \textsc{Hii} regions \citep{Motte18,Kalcheva18}, and in the higher end by estimates of a few Myr for the coexistence of massive stars and molecular clouds before dispersal \citep{VZ2018,Chevance20}.

\subsection{Synthetic broadband images}
\label{subsec:synthetic_broadband_images}
\begin{figure*}
    \centering
    \includegraphics[height=0.3\textheight, width =  \textwidth]{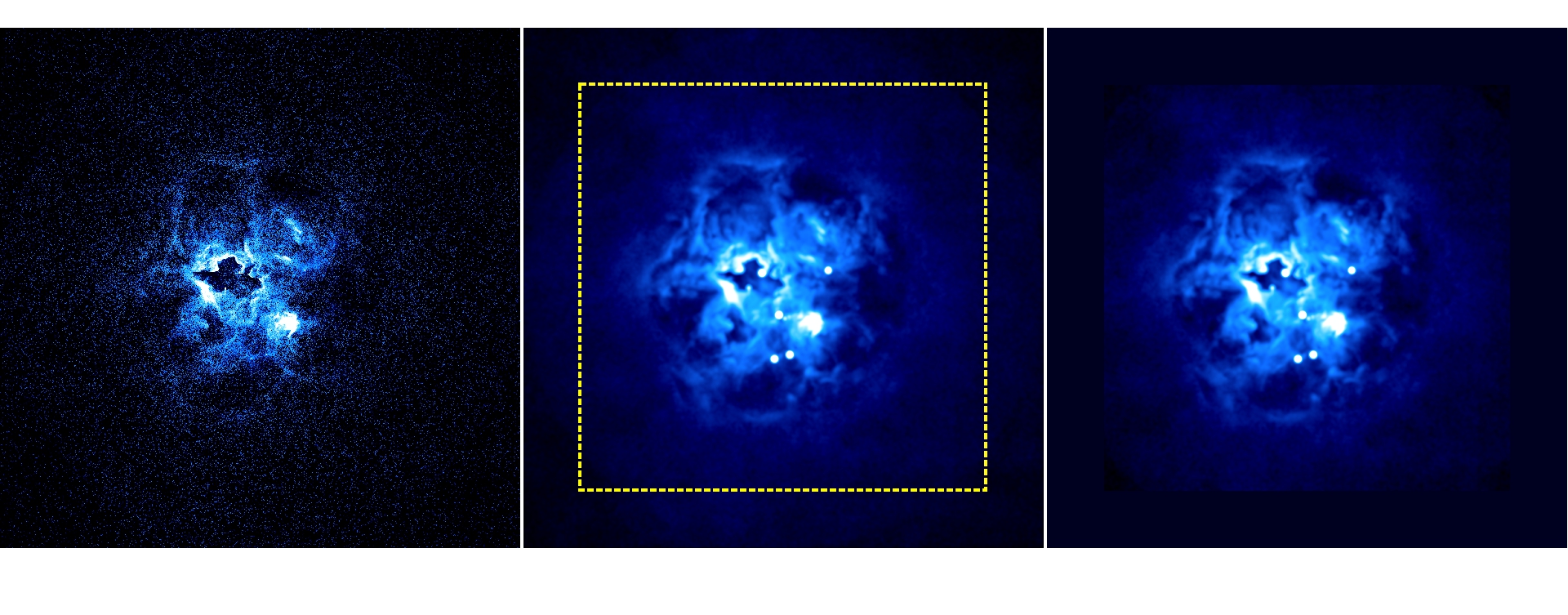}
    \caption{Sketch of the process to obtain a background-subtracted \textit{synthetic observation}. The {\it left} panel shows a raw \texttt{SKIRT} output image. 
    The {\it middle} panel shows an {\it  intermediate synthetic observation} after convolution with the PACS $70~\mathrm{\mu m}$ PSF. A yellow box separates the {\it observed cloud} from the outer region where the {\it background} is defined. The {\it right} panel shows the \textit{synthetic observation} after background subtraction. 
    The color scale is arbitrary to highlight the molecular cloud features. This snapshot corresponds to 3.3 Myr after the first star is born.}
    \label{fig:regions}
\end{figure*}

\texttt{SKIRT} produces idealized synthetic images in selected broadband filters. We simulated observations in the IR, so the broadband used for this work comes principally from \textit{Spitzer} and \textit{Herschel} observatories, and are listed in Table \ref{tab:bandas_usadas}. The left panel of Figure \ref{fig:regions} shows an example of an idealized output image for the PACS $70~\mathrm{\mu m}$ band from the 3.3 Myr snapshot. To include the effects of the spatial instrument response we convolve the \texttt{SKIRT} output images with an approximate point spread function (PSF). We choose 2D Gaussian kernels with full width at half-maximum (FWHM) matched to the PSF of each combination of instrument and band (see Table \ref{tab:bandas_usadas}). We call the resulting image an \textit{intermediate synthetic observation}, exemplified in the central panel of Figure \ref{fig:regions} for the PACS $70~\mathrm{\mu m}$ band. 
%checkV3

We subtract a contribution from the foreground and background emission -- hereafter just called ``background'' for simplicity -- from the intermediate synthetic observations. This background is the sum of dust emission outside of $100~\mathrm{pc}$ central cube plus artificial background emission due to the geometry of our implementation of the ISRF.
The latter artificial background comes from the ISRF photons that reach the synthetic camera without interacting with any dust in the grid (see Section \ref{sec:ISRF}). To define the background in each of the intermediate synthetic observations, we choose a central sub-region of $68.75 \times 68.75 ~ \mathrm{arcmin}^2$ from the total $88 \times 88 ~ \mathrm{arcmin}^2$ image. This sub-region is called the \textit{observed cloud} and is represented as a yellow box in the central panel of Figure \ref{fig:regions}. The background is then defined as the average intensity outside the observed cloud and is subtracted from the flux of the intermediate synthetic observation. We call this background-subtracted sub-image the \textit{synthetic observation}, as shown in the right panel of Figure \ref{fig:regions}. 
These are used to extract photometry and derive the physical properties of the simulated clouds in such a way to mimic observational methods.

%%%chech V1

%FWHM table
\begin{table}%[t]
    \caption{Photometric bands used to create the synthetic observations from the radiative transfer simulations. Taken from \cite{de2014high}.}%, fazio2004infrared}.}
    \begin{center}
        \begin{tabular}{ l  c  c }
            \dtoprule
            Band & $\lambda$ & FWHM  \\ 
             & [$\mu \rm m$] & [arcsec]  \\ \hline
            %GALEX FUV & 0.15 & 4.2 \\
            IRAC 3.6 & 3.6 & 1.7  \\
            IRAC 4.5 &  3.6 & 1.7 \\ 
            IRAC 5.8 & 5.8 & 1.9  \\
            IRAC 8.0 & 8.0 & 2.0\\
            WISE 3 & 12 & 6.5 \\
            MIPS 24 & 24 & 6 \\ 
            PACS 70 & 70 & 5.8 \\
            PACS 100 & 100 & 6.8 \\
            PACS 160 & 160 & 12.1 \\
            SPIRE 250 & 250 & 18.2 \\
            SPIRE 350 & 350 & 24.9 \\
            SPIRE 500 & 500 & 36.3 \\
            \dbottomrule
        \end{tabular}
    \label{tab:bandas_usadas}
    \end{center}
\end{table}
%%%%%%%%%%%%%%%%%%%%%% end Synthetic observation
%%%%%%%%%%%%%%%%%%%%%%%%%%%%%%%%% begin Analysis
\section{Analysis}
\label{sec:results}
The advantage of dealing with numerical simulations is that their physical properties are explicitly known. Therefore, we can apply standard analysis techniques to the synthetic observations and check their ability in recovering properties such as the dust mass and temperature. 
In this section we perform an analysis of the synthetic observations: we measure the far-IR fluxes and fit them with a modified blackbody model, then derive the corresponding dust mass and temperature. We perform the analysis both for the full cloud fluxes and in a  spatially-resolved manner. 

\subsection{Observed synthetic SEDs and comparisons to observations}
\label{subsec:Observed_SED_comparision_observations}
\begin{figure*}
    \centering
    \includegraphics[width =  \textwidth]{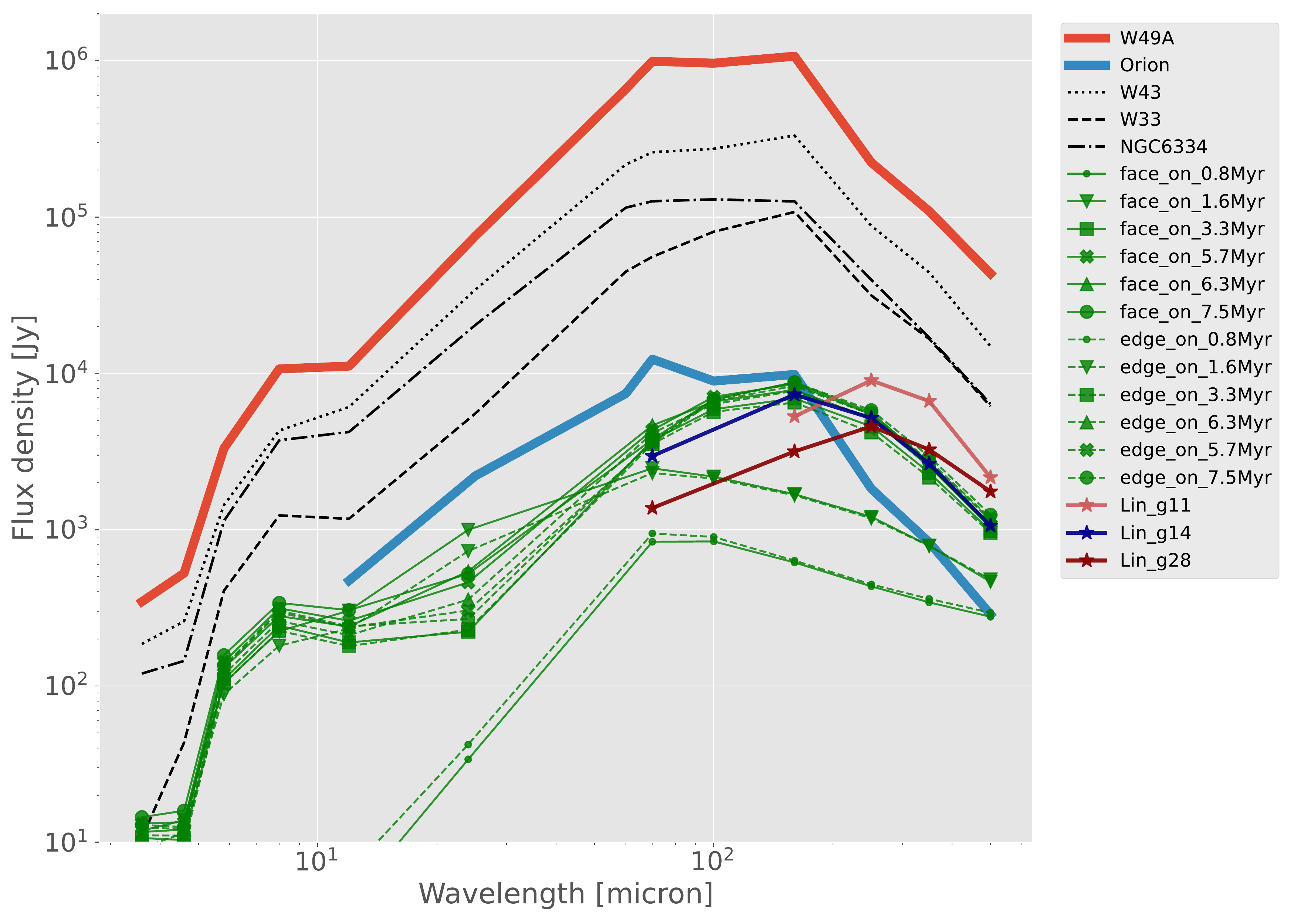}
    \caption{Comparison between the observed synthetic SEDs and those of Galactic star formation regions.
    Green lines show the synthetic SEDs for the face-on (solid lines) and edge-on (dashed) model views. Timesteps in the models use different markers.  
    The lines without markers are SEDs from active clouds in the sample of \cite{binder2018multiwavelength},  while the colored lines with star markers are SEDs from the IRDC sample of \cite{lin2017cloud}. All the fluxes have been scaled to a distance of 2.9 kpc.}
    \label{fig:binder_skirt_yuxin}
\end{figure*}

For each snapshot of the hydrodynamical simulation, we have produced \textit{synthetic observations} in the 12 IR bands listed in Table \ref{tab:bandas_usadas}. We extract the fluxes from these using the previously defined instrumental aperture of 100 pc per side (see Section \ref{subsec:synthetic_broadband_images}). Hence, for each time step, we can construct the respective \textit {observed synthetic} SED. Figure \ref{fig:binder_skirt_yuxin} shows these from mid- to far-IR for the edge-on and face-on line of sights. 
%check V0

Figure \ref{fig:binder_skirt_yuxin} also shows the observed SEDs of a sample of active massive star forming clouds taken from \citet{binder2018multiwavelength}, as well as a sample of 
infrared dark clouds (IRDCs) presented by \citet{lin2017cloud}. All the SEDs were scaled to the average distance ($2.9 ~ \mathrm{kpc}$) of the Galactic sample to enable a more direct comparison.  
%check V0 
The observed synthetic SEDs follow the same time evolution described for the raw SEDs in Section \ref{subsec:Observed_SED_comparision_observations}, but with coarser detail. 
Once the ISRF is removed via background subtraction, we can assume that the emission at these wavelengths is purely from the star formation region. 

In the first time step at 0.8 Myr, when the stellar radiation from the relatively low mass stellar population is still deeply embedded, the peak of the SED is at $\sim 70 ~ \mathrm{\mu m}$, and the emission is dominated by warm dust (at $\geq 40$ K) that is very close to the newly formed stars. 
% explanation of the second step.
In the second time step, at 1.6 Myr, the cloud luminosity at $24 ~ \mathrm{\mu m}$ becomes relevant, even slightly larger than at later times in this band. This represents the transition from dense cores of molecular gas to small \textsc{Hii} regions. Hence, in this phase we observe significant mid-IR emission caused by dust around the young \textsc{Hii} regions that start to expand \citep[e.g.,][]{Deharveng2010,Anderson2014}. The peak of emission is still around $\sim 70 ~ \mathrm{\mu m}$. 

One would expect the synthetic observations of the two earlier snapshots to resemble the SEDs of young objects such as IRDCs, where the emission is dominated by cold dust peaking around $\sim 250 ~ \mu$m  \citep[e.g.,][]{Pillai11,Battersby14,lin2017cloud}. Instead, the shape of those SEDs does not resemble any observed Galactic region. We believe this to be an artifact of the models,  since the young ($\leq 1$ Myr) sinks are populated with main sequence stars rather than protostars, which will have a colder photospheric spectrum \citep{hosokawa2009}.

% check V2
%%% The rest of the snapshots.
Once the material around the massive stars is swept away by the ionizing feedback, UV-optical photons can travel further and heat the dust in the entire simulation grid. This is why the peak of the emission for these synthetic SEDs is around $\sim 160 ~ \mu$m, consistent with the Galactic sample. Also, the sinks in these snapshots are old enough to ensure that they are correctly populated with stars already in the main sequence.
%checkV2

%%% describe the MBB
\subsection{Modified blackbody fits to the SEDs}
\label{integrated_MBBfit}
Modified black body (MBB) fitting using far-IR images is widely used to derive integrated dust properties such as mass, temperature, and emissivity index. % \citep[e.g.,][]{2014A&A...568A..98A,Ladjelate2020,2022MNRAS.510..658P}.
We now apply these classical methods to the synthetic photometry of the simulated cloud.
To do so, the far IR \textit{Herschel} bands were fitted using MBB (either 1 or 2 components) under the assumption of optically thin dust emission. The general form of a MBB is given by: 
\begin{equation}
 S_\nu = \frac{M_\mathrm{dust}}{d^2} \kappa_\nu B_\nu(T_\mathrm{dust}),
 \end{equation}
where $M_\mathrm{dust}$ is the dust mass, $\kappa_\nu$ is the dust opacity, $d$ is the distance to the cloud, and $B_\nu(T_\mathrm{dust})$ is the Planck function at a dust temperature $T_\mathrm{dust}$.

\begin{comment}
defined as 
\begin{equation}
    B_{\nu}(T_{dust}) = \frac{2h\nu^3}{c^2} \frac{1}{c^{h\nu / \kappa_B T_{dust}} - 1}
\end{equation}
where $h$ is the Planck constant, $c$ is the speed of light, and $\kappa_B$ is the Boltzmann constant. 
\end{comment}

The frequency-dependent dust opacity was assumed to follow a power-law of the form: 
\begin{equation}
    \kappa_\nu = \kappa_0 \times \left( \frac{\nu}{\nu_0} \right) ^ \beta,
    \label{eq:kappa_nu}
\end{equation}
where $\beta$ is the dust emissivity index. 

We use the same dust optical properties as in the radiative transfer simulations \citep{draine2007infrared}. From a fit to the far-IR opacities, we obtain an emissivity index $\beta = 2.08 $ and an opacity at $350 ~ \mathrm{\mu m}$ $\kappa_0 = \kappa_{350} = 2.58 ~ \mathrm{ cm^2 ~ g^{-1}}$, which will be kept constant in the rest of this work. Therefore, the only free parameters in the MBB fit will be the dust mass and temperature. The fit was performed with the python module \texttt{lmfit}  \citep{matt_newville_2021_5570790}.

As we have mentioned in Section \ref{sec:synthetic_observations}, the SEDs of the first two timesteps at 0.8 and 1.6 Myr do not have the characteristic modified blackbody shape and are in fact rather flat. For this reason, no acceptable MBB fits could be obtained for these two snapshots. For the rest of the timesteps, the SED shape is very close to the typical MBB  emission, therefore the simulated data are well represented by this model.
%single MBB
First, we performed the fit using a single temperature MBB to calculate the properties of the cloud using the $160$ to $500~\mu$m bands. This approach is frequently used where the observations at 70 and 100 microns are not available \citep[e.g.,][]{2014A&A...568A..98A,Ladjelate2020,2022MNRAS.510..658P}.
The resulting mass and temperature with this model are shown in the right part of Table \ref{tab:fits_results}.

%% two MBB
We also used the slightly more general approach of including data points at 70 and 100 $\mu$m and performing the photometry fitting. %on a pixel-by-pixel basis. 
This allows to recover dust mass in a second (warmer) component at $\sim 30$ K, in addition to the colder component at $\sim 15$ K. The resulting total dust masses following this resolved 2-temperature MBB approach are listed in the left part of the Table \ref{tab:fits_results}, for both lines of sight.  
% check v2
The dust masses derived in this way are systematically larger compared to those obtained using a single MBB.
This is due to the fact that, in the 2-temperature model, the temperature of the colder component is lower than the dust temperature of the single-component MBB. Therefore, a larger cold-dust mass is needed to reach the observed flux levels. 
%chevk V2

%  fit results  table
\begin{table*}[]
    \footnotesize
    \centering
    \begin{threeparttable}[b]
    \caption{Masses and temperatures calculated from the modified blackbody fits for the two lines of sight. The second column shows the mass in the $100 ~ \mathrm{pc}$ cubic central box calculated directly from the dust simulation dust grid. 
    }
    \setlength{\tabcolsep}{4.3 pt} % for the horizontal padding
        \begin{tabular*}{\textwidth}{@{}cc|ccc|ccc|cc|cc@{}}
            \dtoprule
             \multicolumn{2}{c|}{} & \multicolumn{6}{c|}{Two modified BB} & \multicolumn{4}{c}{Single modified BB} \\
            \hline %\\
            \multicolumn{2}{c|}{Intrinsic} & \multicolumn{3}{c|}{face-on} & \multicolumn{3}{c|}{edge-on} & \multicolumn{2}{c|}{face-on} & \multicolumn{2}{c}{edge-on}\\ \hline
            Snapshot & $M_\mathrm{100pc}$ &  mass & $T_\mathrm{warm}$& $T_\mathrm{cold}$ &  mass & $T_\mathrm{warm}$ & $T_\mathrm{cold}$ &  mass & T & mass & T \\
            Myr & $10^2~ M_\odot$ & $M_\odot$ & K & K  & $M_\odot$ & K & K & $M_\odot$ & K & $M_\odot$ & K \\
            \hline
            0.8\tnote{a} & 8.89 & --- & --- & --- &  --- & --- & --- & --- & ---& --- & ---\\ 
            1.6\tnote{a} & 8.93 & --- & --- & --- & --- & --- & --- & --- & ---& --- & ---\\ 
            3.3 & 8.92 & $527 \pm 74$ & $29.8\pm 2.3$ & $15.1 \pm 0.9$ & $531 \pm 80$  & $28.8 \pm 1.8$ & $14.6 \pm 0.9$ & $422 \pm 20$ & $17.1\pm0.2$ & $409 \pm 30$& $16.9 \pm 0.3$ \\ 
            5.7 & 9.01 & $512 \pm 68$ & $30.8\pm 3.3$ & $16.1 \pm 1.0$ & $635 \pm 93$  & $27.3 \pm 1.9$ & $14.9 \pm 1.0$ & $433 \pm 16$ &$17.7\pm0.2$ & $508 \pm 29$& $17.0 \pm 0.3$ \\ 
            6.3 & 9.00 & $506 \pm 66$ & $33.6\pm 3.6$ & $16.0\pm 0.9$&  $626 \pm 85$  & $30.5 \pm 2.4$ & $15.1 \pm 0.8$ & $430 \pm 15$ &$17.4\pm0.2$ & $514 \pm 30$& $16.7 \pm 0.3$\\ 
            7.5 & 9.10 & $500 \pm 65$ & $30.1\pm3.8$ & $16.4\pm1.1 $ &   $655 \pm 92$  & $28.0 \pm 2.3$ & $15.2 \pm 1.0$ & $432 \pm 14$ &$17.8\pm0.2$ & $541 \pm 28$& $17.0 \pm 0.2$\\ 
        \dbottomrule
        \end{tabular*}
    \begin{tablenotes}
        \item [a] Values are not reported because the fit is not reliable.
    \end{tablenotes}
\label{tab:fits_results}
\end{threeparttable}
\end{table*}
%check v3

In both types of fitting, the derived masses for the last 3 snapshots are larger for the edge-on view than for face-on. This is because in the face-on view the warmer dust in the immediate vicinity of the radiating stars is directly seen. Indeed, for these snapshots we observe that fluxes for the face-on view are larger at wavelengths $<160~\mu$m, as compared to those for the edge-on view. Therefore, the global temperature obtained by the MBB fit is slightly larger in the face-on view (see Table \ref{tab:fits_results}), and a slightly lower mass is needed to fit the synthetic photometry.

\subsection{Comparison to intrinsic dust masses} 
\label{sec:intrinsic_mass}
In the previous section we have applied standard analysis techniques such a modified black body fitting, to check their ability in recovering dust properties such as mass and temperature. We have found that this method systematically underestimates the ``real'' mass by at least $30\%$ in the best case scenario (see Table \ref{tab:fits_results}). 

With the aim of analyzing where the differences come from, we have calculated the cumulative mass as a function of temperature from the cells inside the \texttt{SKIRT} dust grid to see what is the amount of dust mass that is too cold to provide a significant contribution in the far-IR \textit{Herschel} bands.
%checkV2
To be consistent with the definition of \textit{observed cloud} from Section \ref{subsec:synthetic_broadband_images}, we use the radiative transfer simulation grid to calculate the mass of dust inside a cube of $100 ~ \mathrm{pc}$ length, that we will dub as $M_\mathrm{100pc}$.
%checkV2
In the second column of Table \ref{tab:fits_results} we show $M_\mathrm{100pc}$ for the different snapshots.
%checkV3

\begin{figure}
    \centering
    \includegraphics[width = 0.45\textwidth]{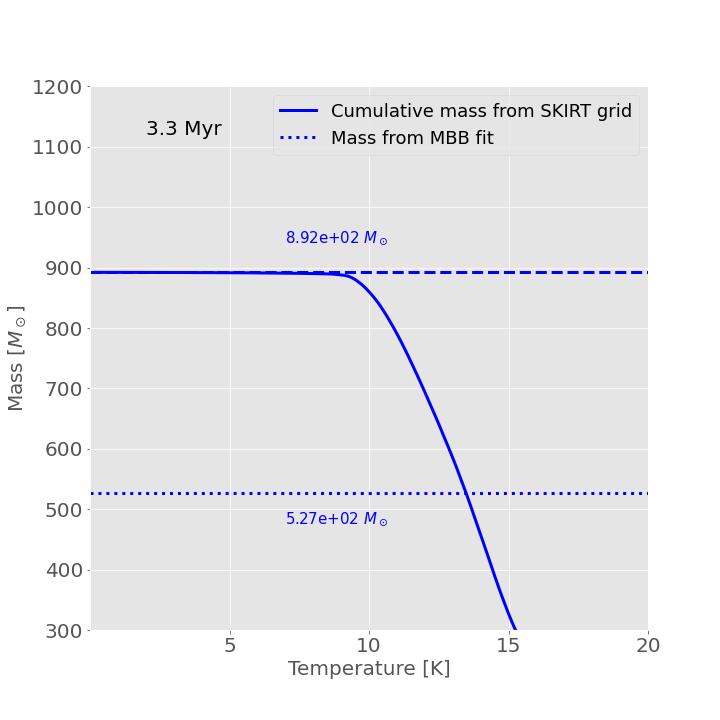}
    \caption{Cumulative dust mass in the \texttt{SKIRT} grid for the 3.3 Myr snapshot, within a box of $M_\mathrm{100pc}$  (solid line). The total mass within this grid (horizontal dashed line) and the total recovered mass from the 2-MBB fits (horizontal dotted line) are also shown for comparison. The intersection between fitted and intrinsic mass occurs at 13.5 K.}
    \label{fig:cumulative_mass}
\end{figure}

Figure \ref{fig:cumulative_mass} shows the cumulative mass function for cells with temperature below a given threshold, for the case of the 3.3 Myr snapshot. The total dust mass recovered from the 2-MBB fits is also shown. The intersection occurs at about 13 K, showing that dust with temperature lower than this value will have an increasingly smaller contribution to the observed far-IR flux.

To be able to estimate the contribution to the far-IR flux of the coldest cells, we create a toy model where each cell inside the $M_\mathrm{100pc}$ cube emits as a modified blackbody with mass and temperature given by the properties of the cell. We define as ``cold'' cells those with $T \leq 15$ K, and as ``warm'' those with $T > 15$ K. 
Then, for each of the two components, we calculate both their dust mass and far-IR fluxes from $70$ to $500 \mu$m.
In Table \ref{tab:mass_vs_flux} we show the results of this experiment as contribution percentages of mass and bolometric far-IR flux for both the cold and warm components. We can see that the bolometric flux is dominated by the warm component, which contributes with $\sim 90 \% $ of the total emission at each snapshot. The cold component, whose mass is around $37-63 \%$ of the total, only contributes with $\sim 10\%$ of the flux. 

\begin{table}[]
    \caption{Percentage of contribution to the total mass and far-IR flux for the ``cold'' ($T \leq 15$ K) and ``warm'' ($T > 15$ K) dust cells.}
    \centering
    \begin{tabular}{c c c c c}
    \dtoprule
       Time & Cold & Cold & Warm & Warm\\
        & mass & flux & mass & flux \\
        Myr & $\%$ & $\%$ & $\%$ & $\%$ \\
        \hline
        0.8 & 99.57 & 7.44 & 0.43 & 92.56 \\
        1.6 & 96.20 & 13.40 & 3.80  & 86.30 \\
        3.3 & 63.33 & 11.90 & 36.67 & 88.10 \\
        5.7 & 47.03 & 9.32  & 52.97 & 90.68 \\
        6.3 & 47.68 & 10.21 & 52.32 & 89.79 \\
        7.5 & 37.72 & 8.21  & 62.28 & 91.79 \\
        \dbottomrule
    \end{tabular}
    \label{tab:mass_vs_flux}
\end{table}

\begin{figure*}
    \centering
    \includegraphics[height =0.33 \textheight]{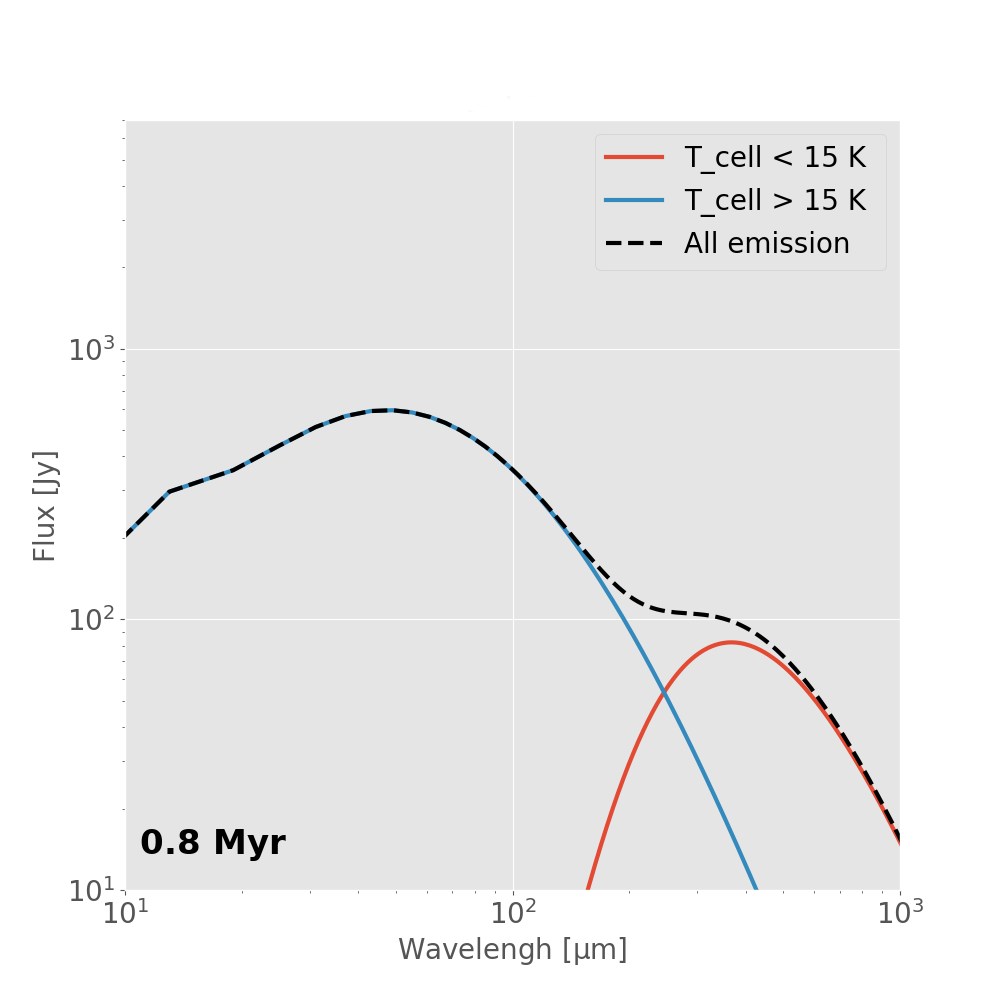}
    \includegraphics[height =0.33 \textheight]{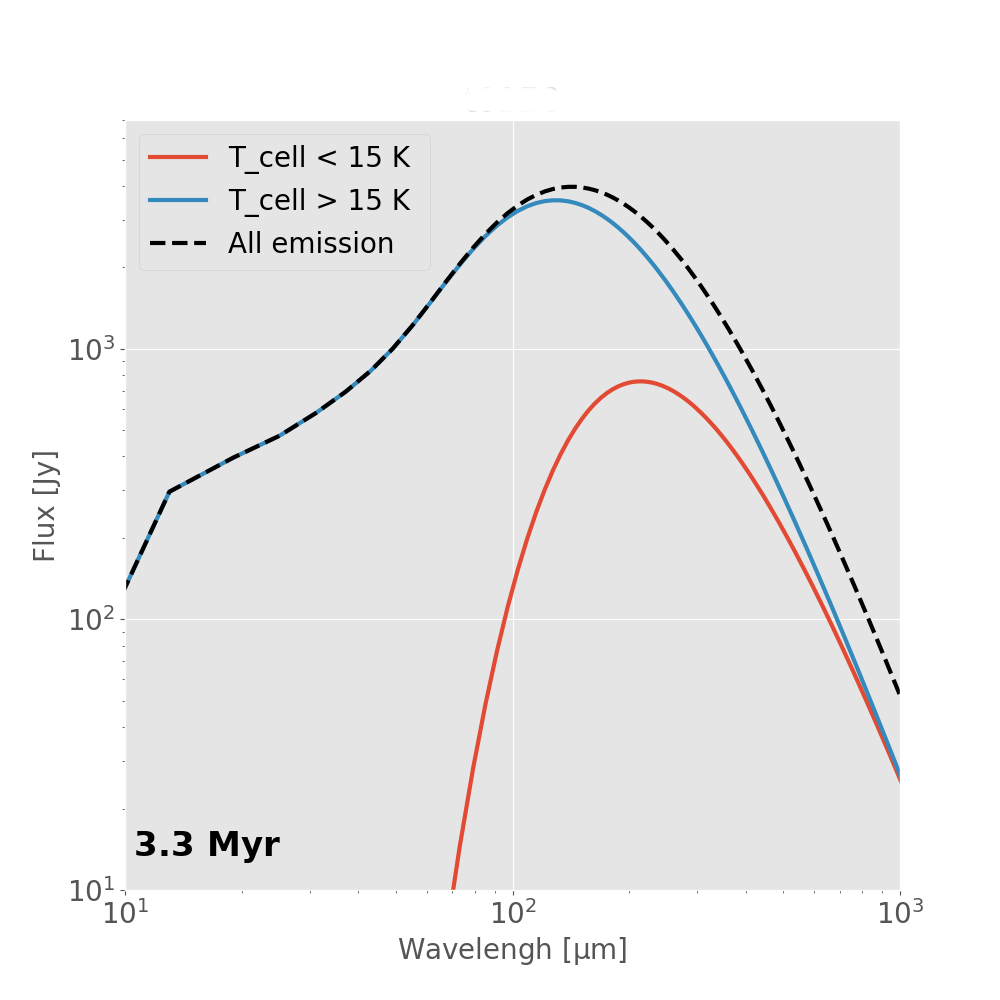}
    \caption{Resulting SEDs for the toy model described in Section \ref{sec:intrinsic_mass}. The blue and red lines show the SEDs of the ``warm'' and ``cold'' components, respectively. The black dashed line is the sum of the two components. The left panel shows the earliest snapshot at 0.8 Myr. The right panel corresponds to a more evolved snapshot at 3.3 Myr.}
    \label{fig:toy_model_SED}
\end{figure*}

In Figure \ref{fig:toy_model_SED} we show the SEDs of these two components, obtained as the sum of the MBB models for each cell. In the youngest snapshot, the emission of the warm component dominates the IR SED at all wavelengths $< 250$ $\mu$m, where the cold dust emission kicks in. Interestingly enough, at the later time steps, it is the warm dust that dominates the emission up to at the longest wavelengths. It is also worth noting that the difference in wavelength of the two emission peaks decreases monotonically with time, probably due to a more homogeneous dust temperature distribution. The comparable luminosities of the two components at  0.8 and 1.6 Myr also explains why the SEDs are very flat in these snapshots.

In summary, the presence of very cold cells that are underluminous in the far-IR explains the difference between the ``real'' mass in the hydrodynamical simulation and the one derived from MBB fits.  We consider this to be a real effect that could bias observations. 

\subsection{Pixel by pixel fit}
\label{pix_by_pix}
An advantage of dealing with spatially resolved data is that it is possible to perform pixel-by-pixel MBB fitting. One can then obtain maps of the dust surface density and temperature for the region under scrutiny.
In unresolved observations the properties that we recover are inevitably luminosity-weighted ones, therefore a clear advantage of resolved analysis is that we are able to follow the morphological evolution of these quantities by analyzing changes in the synthetic maps of the different timesteps of the hydrodynamical simulation.
%\footnote{Note that it is not trivial to get, directly from the RT, the dust average temperature within a pixel. On the other hand, this analysis, while not perfect, can provide a fair estimate of this.}.

In order to perform this analysis, we convolve all the \textit{synthetic observations} to the  $500 ~ \mu$m \textit{Herschel} PSF, to match them in angular resolution. To optimize the calculations, we regrid the images in such a way that one pixel roughly corresponds to 1/3 of the PSF width. Like this, we go from a pixel scale of $1.32"$ to $12.1"$. Since in the construction of the synthetic observations we performed a background subtraction, it is possible that some pixels will end up having very low or even negative fluxes. Hence, we only consider pixels with flux density $> 1$ mJy to ensure all bands can be used. Furthermore, when analyzing the results, we discard those pixels with $\chi^2 \geq 10$, for which dust properties are likely not well constrained.  

\begin{figure*}
    \centering
    \includegraphics[width=\textwidth,height=0.5\textheight,trim={2cm 2cm 0cm 0cm},clip]{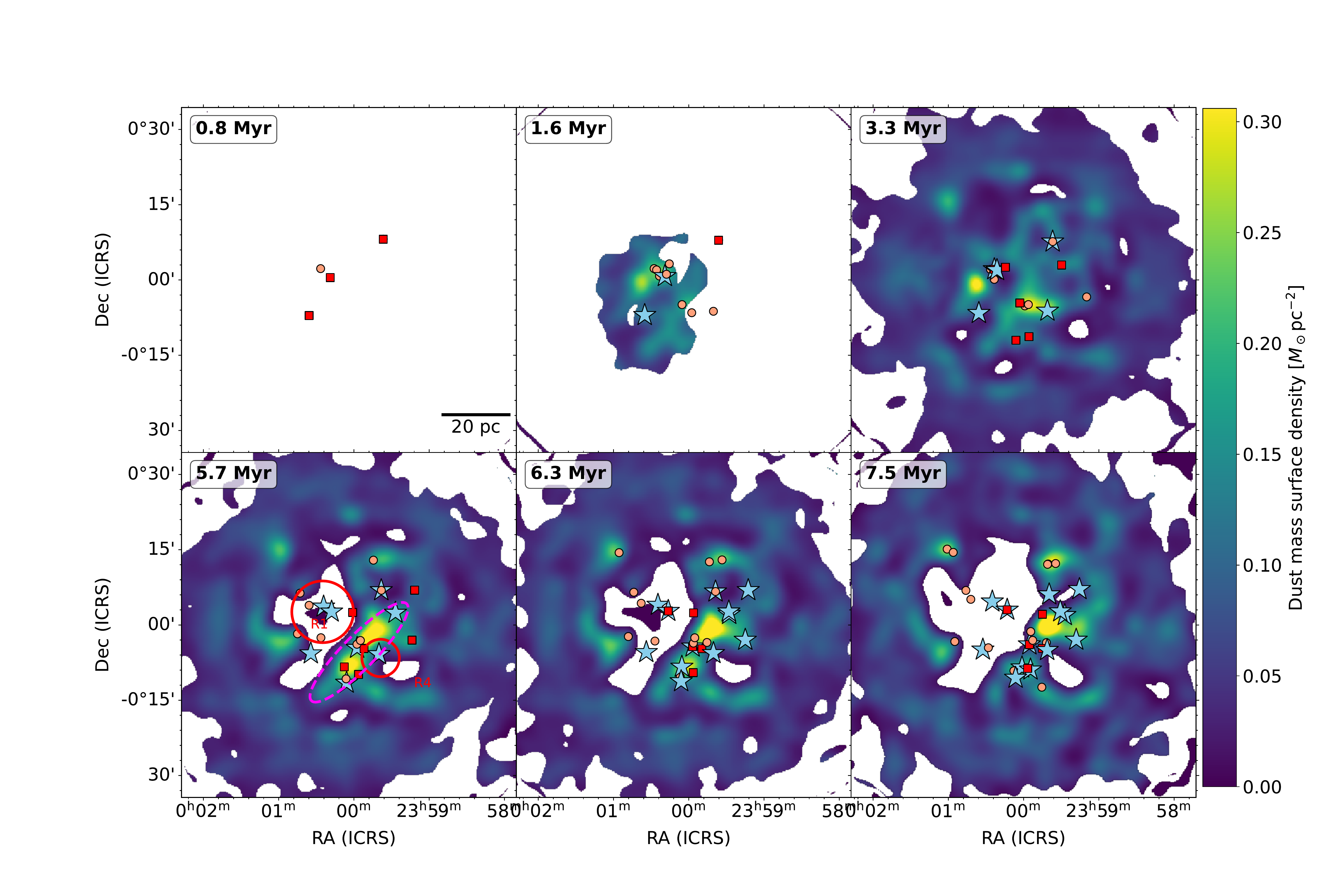}
    \caption{Dust surface density maps  ($M_\odot ~ \mathrm{pc^{-2}}$) resulting from the spatially resolved fitting to the face-on synthetic observations at various timesteps. The markers represent the projected sink positions, where blue stars are used for sinks with masses $\geq 100~ M_\odot$, red squares for sinks with masses between $ 50~ M_\odot$ and $100~ M_\odot$, and orange circles the sinks with $ \leq 50~ M_\odot$. In the bottom left panel, the magenta dashed ellipse represents the overdensity formed by the expansion of the two \textsc{Hii} regions marked by red circles. 
    Pixels that do not meet the conditions for a good fit are blanked.}
    \label{fig:mosaico_face_total_mass}
\end{figure*}

\begin{figure*}[!t]
    \centering
    \includegraphics[width=\textwidth,height=0.5\textheight,trim={2cm 2cm 0cm 0cm},clip]{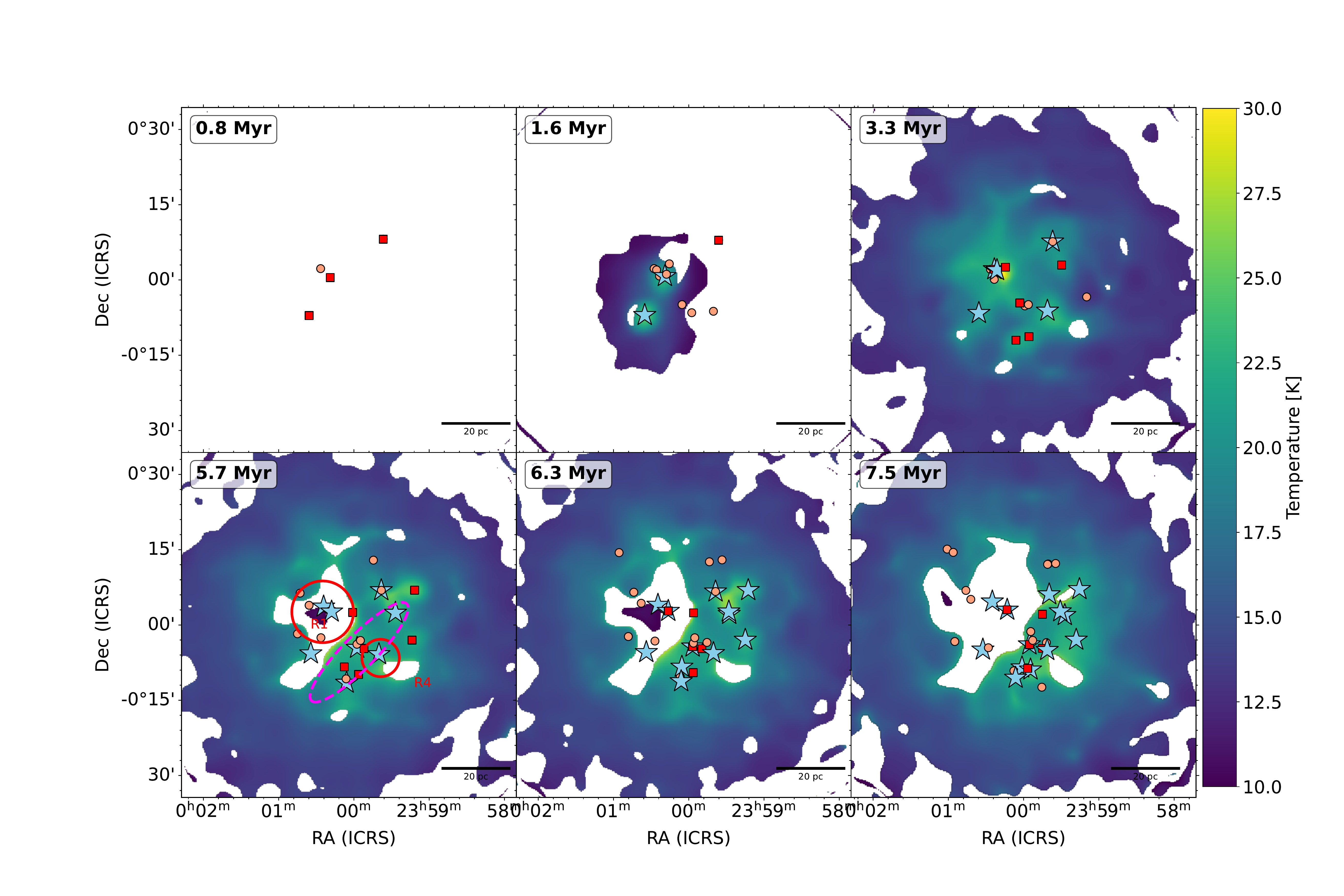}
    \caption{Dust temperature maps in Kelvin resulting from the spatially resolved fitting to the face-on synthetic observations at various timesteps.  Labels and symbols are the same as in Figure \ref{fig:mosaico_face_total_mass}.}
    %The markers represent the projected sinks positions, in blue stars for the sinks with masses $\geq 100~ M_\odot$, in red squares for sinks with masses between $ 50~ M_\odot$ and $100~ M_\odot$, and in orange circles the sinks with  $ \leq 50~ M_\odot$.
    %In the bottom left panel the magenta dashed ellipse represents the overdensity formed by the expansion of the two \textsc{Hii} regions marked by red circles. Pixels that do not meet the conditions for a good fit are blanked.}
    \label{fig:mosaico_face_total_temperature}
\end{figure*}

We use a single temperature MBB to reproduce emission within the $160-500~\mu$m range. 
The resulting maps of dust surface density and temperature for the face-on line of sight are shown in Figures \ref{fig:mosaico_face_total_mass} and \ref{fig:mosaico_face_total_temperature}, respectively. 
%checkV5

Next, we analyze the surface density maps for the different snapshots. In the first time step (upper-left panel of Figure \ref{fig:mosaico_face_total_mass}) the vast majority of pixels have a poor fit. As mentioned before, the reason is that at wavelengths $\lambda \gtrsim 200 ~ \mu$m their SED is basically flat.  Therefore, the surface density maps could not be determined for the most part.
Figure \ref{fig:bad_pixels_sample} in Appendix \ref{sec:app_A} shows an example SED of a pixel with a bad fit due to the above-mentioned issue.

As the cloud evolves to 1.6 and 3.3 Myr (top center and right panels in Figure \ref{fig:mosaico_face_total_mass}), we see that more sinks start to appear around the central massive one. Even though the newly formed sinks are of low mass ($<50~M_\odot$), the additional luminosity that they inject is enough to heat dust over more extended areas, and at 3.3 Myr this heated dust already appears in the recovered maps over almost the entire cloud. We note that in this snapshot the surface density around the most massive sink in the center has already started to decrease, as a consequence of the first appearance of its corresponding \textsc{Hii} region. 

Since the dust density in our radiative transfer models is proportional to the complement of ionized gas in each cell, the dust disappears once the gas has been completely ionized by the massive stars, and thus the model \textsc{Hii} regions have no IR emission. This is why dust surface density cavities become evident in the lower panels of Figure \ref{fig:mosaico_face_total_mass}. 
Another consequence of the formation of the \textsc{Hii} regions is that they push neutral material and create overdensities in their periphery. An example of this is the overdensity or ridge that becomes prominent in the bottom-left panel of Figure \ref{fig:mosaico_face_total_mass}, created by the \textsc{Hii} regions labeled R1 and R4 by \citet{ZA+19}. Over the last two snapshots, at 6.3 and 7.5 Myr, the \textsc{Hii} regions continue their expansion, and it is seen how the sinks that are born within the triggered ridge start to disrupt their parental cloud. 

During the last three snapshots shown in Figure \ref{fig:mosaico_face_total_mass}, the triggered ridge becomes smaller but does not totally disappear. Several effects are in place here. The first is evolutionary. Most of the sinks that are located close to this ridge are born at later stages of the simulation, and further gas dispersal is yet to happen. 
In fact, the same happens to the central region (R1), which contains massive sinks already in the 3.3 Myr snapshot, but develops a prominent cavity until later. Second, the formation of this ridge appears to be due to triggering \citep[e.g.,][]{Deharveng2012} by the collective action of the surrounding sinks. Third, feedback from some of these sinks appears to act more efficiently toward the outward direction of the cloud, where secondary cavities appear. 
And finally, a projection effect. We note how these (column density) cavities are evident in the face-on views (Fig. \ref{fig:mosaico_face_total_mass}) but not when the cloud is viewed edge-on (Fig. \ref{fig:mosaico_edge_total_mass}). 

This description from synthetic observations is consistent with the analysis of \cite{ZA+19}. Note that the morphological evolution described above is much harder to detect and follow if the cloud is viewed edge-on (see Figure \ref{fig:mosaico_edge_total_mass} in Appendix \ref{sec:app_B}). 

We also calculated the total dust masses from the spatially resolved maps using pixel-by-pixel 2-temperature MBB fitting. A comparison of the results is shown in Table \ref{tab:masses_pix_by_pix}. The use of 2 MBB components allows to recover  $\sim 10\%$ more mass than with only 1 MBB, likely because in some single resolved regions the smaller flux contribution of the coldest cells (see Section \ref{sec:intrinsic_mass}) can be recovered.

%checkV1
%%%%%%%%%%% table of mass obtained by the fix-to-fix
\begin{table}[]
\label{pix-by-pix_results}
    \centering
    \begin{tabular}{c c c c}
    \dtoprule
        snapshot & $\mathrm{M_{100pc}}$ & 1 MBB & 2 MBB \\
        Myr & $10^2~ M_\odot$ & $M_\odot$ & $M_\odot$ \\
        \hline
        0.8 & 8.89 & ---   & ---   \\
        1.6 & 8.93 & 142.81 & 602.74 \\
        3.3 & 8.92 & 486.07 & 584.31 \\
        5.7 & 9.01 & 498.49 & 585.61 \\
        6.3 & 9.00 & 496.21 & 576.17 \\
        7.5 & 9.10 & 518.21 & 599.53 \\
    \dtoprule
    \end{tabular}
    \caption{Total masses calculated using pixel-by-pixel MBB fitting for both the single and 2-temperature MBB models in the face-on line of sight.}
    \label{tab:masses_pix_by_pix}
\end{table}

We now describe the dust temperature evolution.  Figure \ref{fig:mosaico_face_total_temperature} shows the inside-out evolution of dust heating by star formation as the cloud evolves in time.
In the snapshot at 3.3 Myr, the stellar radiation field is already strong enough to heat the entire cloud. At this time, $76~\%$ of the pixels show temperatures between 12 and 18 K,  the average temperature being around 14.8 K. The pixels with the highest temperature ($\sim 30.5$ K) are located around the massive star formation sites, and we can see a gradient in temperature from the stars to the edge of the cloud.
%checkV1

In the following snapshots at 5.7, 6.3 and 7.5 Myr (bottom panels of Figure \ref{fig:mosaico_edge_total_temperature}), $79~\%$ of the pixels with valid fits have temperatures between 12 and 18 K, the mean temperature is now $\sim 15.5$ K and the maximum temperature is $\sim 33.3$ K.   

%%%%%%%%%%%%%%%%%%%%%%%%%%%%%%%% SFR
\subsection{Star formation rate}
\label{subsec:SFR_calibrations}
One of the most important physical properties of molecular clouds is the rate at which they form stars. %, and this is inferred from different observable. 
Mathematically, the SFR is defined as the mass of stars formed per unit time:
\begin{equation}
SFR(t) = \frac{\Delta M_{\star}}{\Delta t}.
    \label{eq:SFR_definition}
\end{equation}
On the other hand, hydrodynamic simulations give us the mass in sink particles in discrete time steps instead of continuous ones. Thus, we can only calculate a time-averaged  SFR: 
\begin{equation}
    \langle SFR(t) \rangle = \frac{ M_\star(t)-M_\star(t_0)}{t-t_0}.
    \label{SFR_from_sim}
\end{equation}

One of the biggest problems in estimating the SFR from observations, is that calculating the total mass of newly formed stars is challenging. This can only be done in a few cases where the regions are close enough so that individual stellar objects can be counted \citep[e.g.,][]{Evans09,Gutermuth2009}. Moreover, even in the closest star-forming regions, the youngest YSOs are deeply embedded and the calculation of their (proto)stellar masses is uncertain and depends on the modelling of the effect the emitted radiation on their envelopes \citep[e.g.,][]{Robitaille_2017YSOmodels, DeBuizer2017}.  
%check V1

In an extragalactic context, the SFR is calculated using its relation with the luminosity of various tracers \citep [for a full review of the topic see][]{Kenn_Evans_2012review}, usually given in the form of:
\begin{equation}
 \frac{SFR_\lambda}{M_\odot ~\mathrm{yr^{-1}}} = a_\lambda \left( \frac{\nu L_\nu(\lambda)}{\mathrm{erg ~ s^{-1}}} \right)^b,
 \label{eq:SFR_equation}
\end{equation}
where $b=1$ in almost all the cases, and $a$ is called the calibration coefficient.
%check V1
Two of the most used tracers in the IR domain are the monochromatic luminosities at $24 ~ \mathrm{\mu m}$ \citep{Kenn_Evans_2012review} and $70 ~ \mathrm{\mu m}$ \citep{Li2010ApJ}, for which the calibration coefficients are $a_{24} = 2.03\times10^{-43} $ and $a_{70} = 1.7\times10^{-43}$ respectively. Another tracer that is commonly used is the total IR (TIR) luminosity from $8$ to $1000 ~ \mathrm{\mu m}$  \citep{Murphy2011ApJ}, for which $a_{\mathrm{IR}} = 3.88\times10^{-44}$. The SFR calculated with these estimators is sensitive to different timescales (namely different $\Delta t$ in Equation \ref{eq:SFR_definition}), ranging from $\sim 10$ to $\sim 100$ Myr. Using a sample of Galactic star forming regions for which the more massive components of the stellar population could be resolved, \cite{binder2018multiwavelength} recalculated the values of the $a_\lambda$ coefficients. 

Here we perform a similar analysis, checking both how the calibration of \cite{binder2018multiwavelength} applies to our simulation, and calculating independently the aforementioned calibration coefficients. We note that, on the one hand, using the synthetic observations permit us to know all the intrinsic physical properties of the simulations, but on the other hand, we are only sampling one particular star forming region, and hence cannot explore the diversity of the physical properties that characterize the ensemble of regions found in galaxies. We also note that the latest snapshot of the simulation that we use is at 7.5 Myr after the onset of star formation, which is fairly similar to the time range probed by the 24 $\mu$m luminosity. On the other hand, both the 70 $\mu$m and the bolometric far-IR luminosities are sensitive to larger periods, typically $\sim 100$ Myr, which is significantly larger than the simulated interval. 

\begin{figure}
    \centering
    \includegraphics[width =0.45 \textwidth]{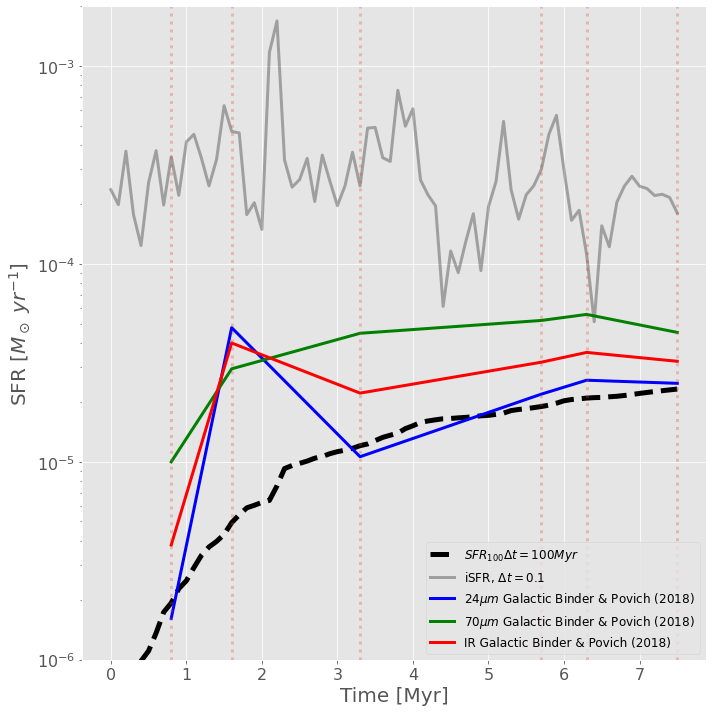}
    \caption{Predicted SFR for the simulation using the Galactic calibration from \cite{binder2018multiwavelength}, using the monochromatic luminosities at 24 (blue line) and 70 (green line) $\mu$m. The red line shows the predicted SFR using the total IR luminosity. 
    The intrinsic iSFR and $\mathrm{SFR_{100}}$ are shown as a solid gray line and dashed black line, respectively.}
    \label{fig:iSFR_raSFR}
\end{figure}

In order to derive the coefficients as in \cite{binder2018multiwavelength}, we first need to calculate the intrinsic SFR of the simulation. \cite{ZA+19} gives the value of mass that is locked in sinks at each snapshot, which are separated by 0.1 Myr. From this, the SFR can be calculated in two ways: 1) the ``instantaneous'' SFR (hereafter iSFR), which is the difference in the mass contained in sinks between two successive simulation snapshots, and 2) the ``100-Myr averaged'' SFR (hereafter $\mathrm{SFR_{100}}$), which is calculated using Equation \ref{SFR_from_sim}. For this, we set the initial time to $t_0 = 92.5$ Myr before the first sink was  born. 
Like this, \cite{binder2018multiwavelength} found values of $a_{24} = 3.7 \pm 2.4 \times 10^{-43}$, $a_{70} = 2.7 \pm 1.4 \times 10^{-43}$ and $a_\mathrm{IR} = 1.2 \pm 0.7 \times 10^{-43}$, for the monochromatic and total IR tracers, respectively. We have used their prescriptions to calculate the predicted SFR of the simulated cloud from our synthetic observations (see Figure \ref{fig:iSFR_raSFR}). 

Conversely, we calculate the $a$ parameters using Equation \ref{eq:SFR_equation} and SFR$_{100}$ as a reference, for the three tracers mentioned above from our synthetics observations. We find that the best values for the calibration coefficients, averaging the snapshots from 3.3 to 7.5 Myr and lines of sight, are $a_{24} = 4.26 \times 10^{-43}$, $a_{70} = 1.08 \times 10^{-43}$ and $a_\mathrm{IR} = 8.24 \times 10^{-44}$. These values are 1.15, 0.4, and 0.69 times the values reported by \cite{binder2018multiwavelength} for the 24, 70, and TIR tracers.

%%%%%%%%%%%%%%%%%%%%%%%%%%%%%%%%%%%%%%%%%%%%%%%%%%%%%%%%%%%%%%
\section{Discussion} \label{sec:discussion}
Creating synthetic observations of simulated star formation regions gives the opportunity to analyze their SED, exploiting standard techniques which are usually applied to real observations, and helping to determine possible biases and limitations of these very same techniques. Furthermore, as we have seen in the previous section, the comparison with observed data can provide important hints to the inevitable limitations carried on by numerical simulations. In this section we give further  interpretation of the results presented in the previous sections. 
\subsection{The role of the interstellar radiation field} 
\label{subsec:discution_isrf}
In contrast to other radiative transfer codes such as \textsc{Hyperion}  \citep{Robitalle2011_hyperion}, where the dust temperature can be started at a specific baseline  temperature during the radiative transfer \citep[see for example, ][]{koepferl17a,koepferl17b,koepferl17c}, in \texttt{SKIRT} the dust temperature is calculated self-consistently by design.
Then, when considering the simulated molecular cloud as an isolated system, we found that the peak of the IR SED, especially in the first evolutionary stages, is located at shorter wavelengths with respect to similar real star forming regions. 

This indicated higher dust temperatures with respect to observations for the parts of the cloud dominating the SED. For the simulated cloud these parts were the pockets of gas around the few sinks formed at early times, but in real IRDCs the bulk of the luminosity comes from the extended cloud itself \citep{lin2017cloud}. 
When we analyzed the dust temperature distribution of the earlier snapshots as obtained from the spatial grid of the RT simulation, we found an average mass-weighed value of about 3.7 K, an extremely low temperature even for IRDCs. 
This turned out to be due to the limited stellar radiation being mostly absorbed ``in situ'', and therefore not being able to reach the outer regions. This motivated us to include the ISRF as an extra source of radiation, which is effective in heating dust over the entire cloud to realistic values in the earliest snapshots equivalent to young clouds such as IRDCs.
The addition of this external radiation component is physically motivated and represents a step further toward a more realistic representation of dust emission star forming clouds. 

In order to assess the importance of the ISRF, we compare RT models where we only use stars as sources of radiation versus models where the ISRF is also included. 
Figure \ref{fig:24_microns_bands_and_SEDS} shows the resulting SEDs and projected column densities for the snapshots at 0.8, 1.6, and 3.3 Myr.  
The surface density maps in Figure \ref{fig:24_microns_bands_and_SEDS} show the presence of important relative voids of material in many lines of sight. These voids are initially formed as a  consequence of molecular cloud collapse, which collects the material in filaments, leaving low-density regions around them \citep[e.g.,][]{Gomez_2014}, and at later times are also due to the expansion of \textsc{Hii} regions \citep{ZA+19}. 
The ISRF can freely cross the cloud through these voids. 
In the earliest snapshot we see that the ISRF dominates the total SED from the UV up to $\sim 10~\mu$m. Thus the above mentioned effect is significant.  
In the following snapshot at 1.6 Myr, the ISRF direct emission is now overwhelmed by emission from the cloud itself. At this age, the stellar radiation begins to dominate dust heating, even in the outer regions of the cloud. 
The ISRF contributes significantly to the SED at $\sim 2~\mu$m and $> 500~\mu$m.
At 3.3 Myr the relative contribution of the ISRF to the SED is even smaller.

Finally, we note that even though the ISRF was included to pre-heat the cloud dust to realistic values prior to star formation, the effect here discussed does not affect the analysis in Section \ref{sec:results}, because it is removed as part of the background subtraction to the create the observed synthetic SEDs.

{\captionsetup[subfigure]{labelformat=empty}
\begin{figure*}
    %\centering
    \subfloat[]{\includegraphics[width=0.33\textwidth,height=0.23\textheight]{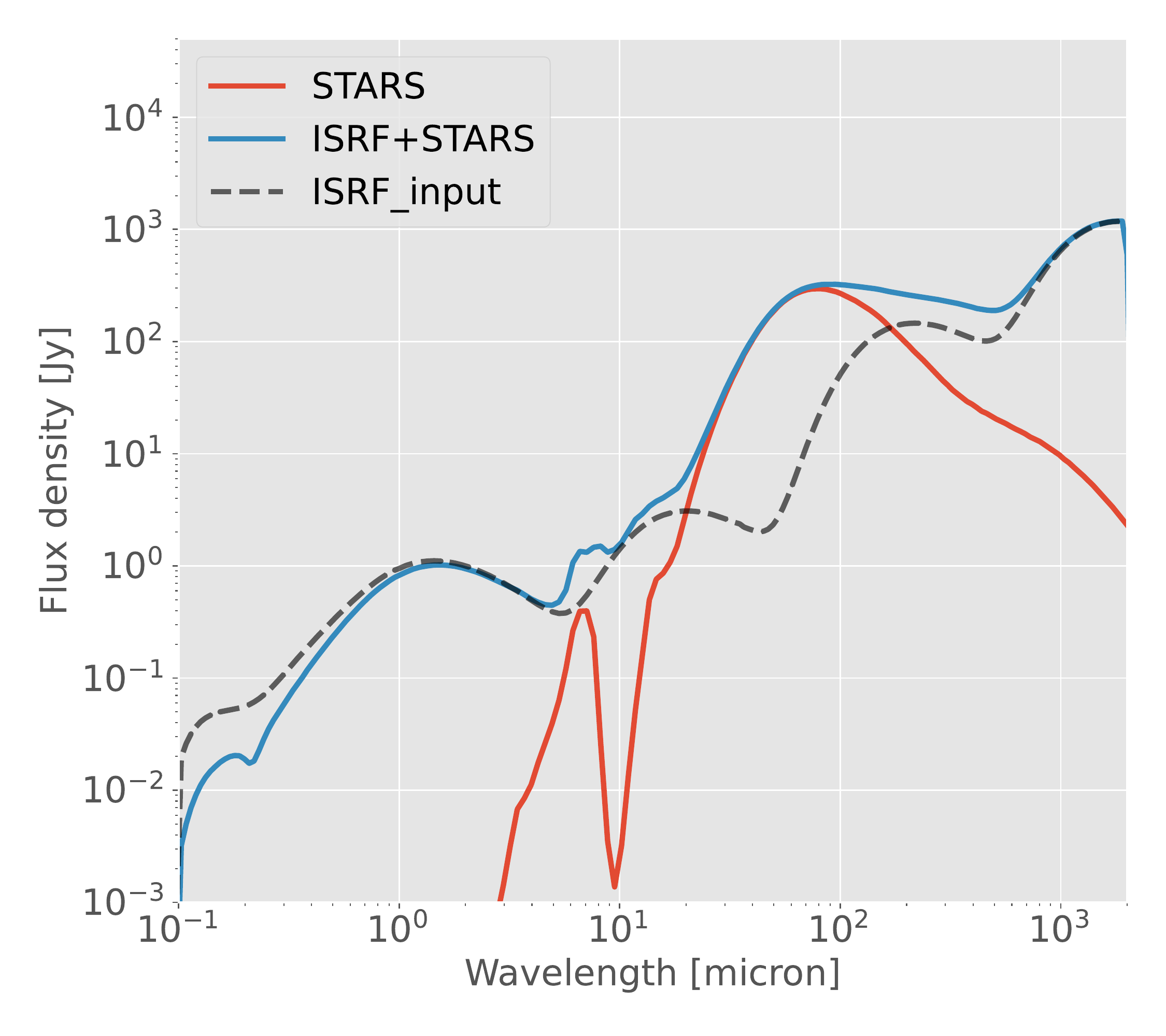}}
    \subfloat[]{\includegraphics[width=0.33\textwidth,height=0.23\textheight]{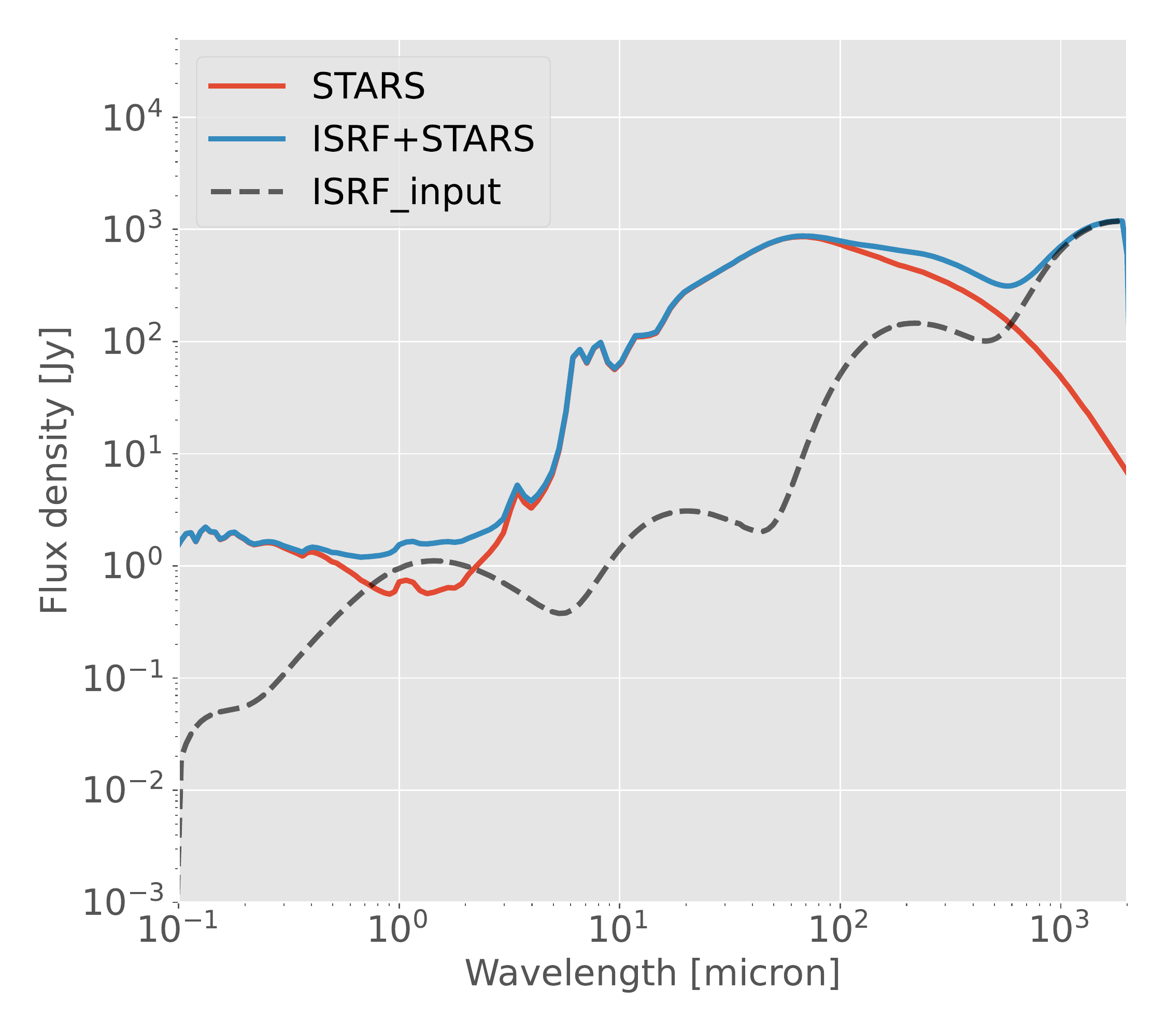}}
    \subfloat[]{\includegraphics[width=0.33\textwidth,height=0.23\textheight]{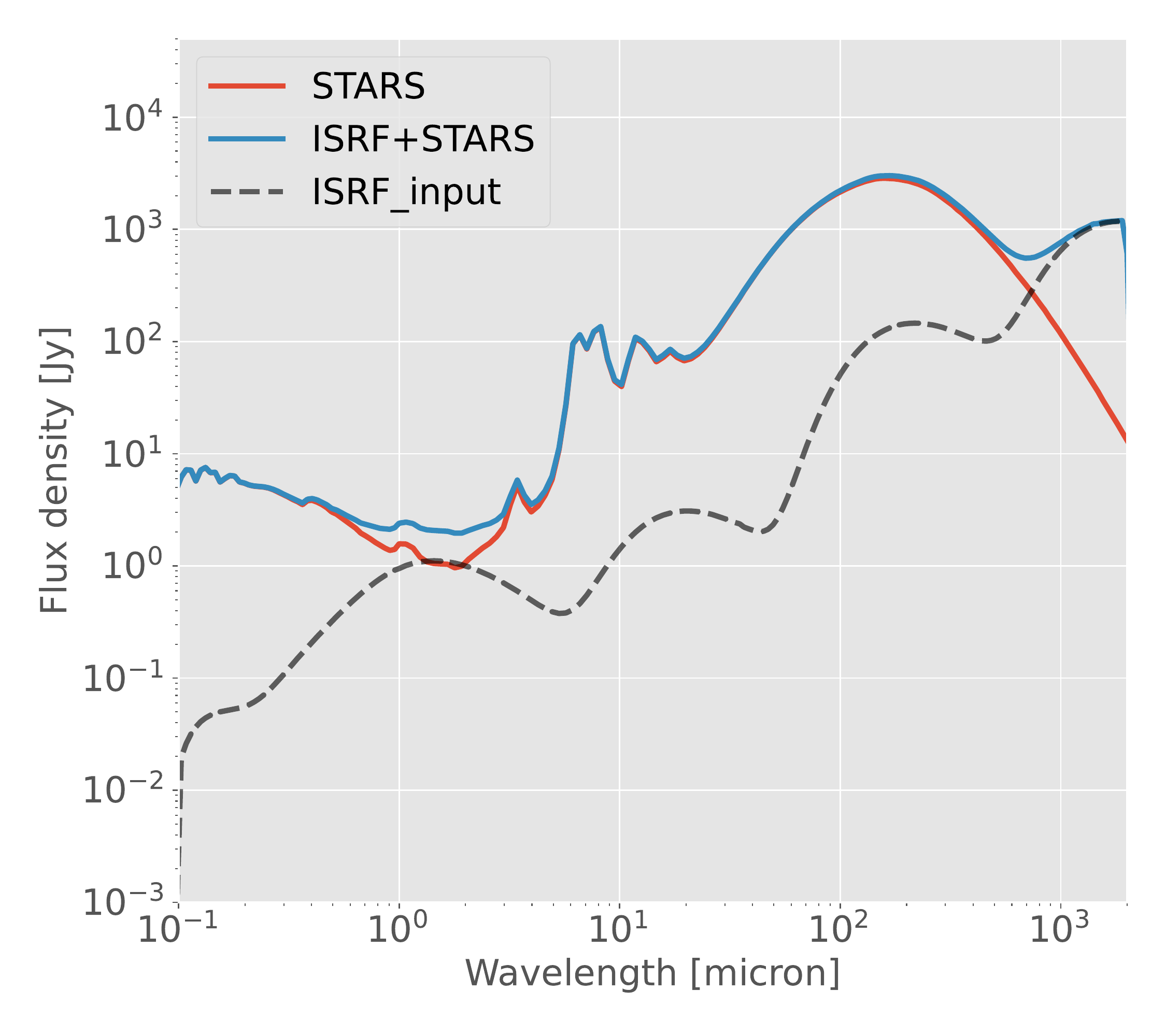}}\\
    \subfloat[]{\includegraphics[width=0.33\textwidth,height=0.23\textheight,trim={0cm 2cm 0cm 0cm},clip]{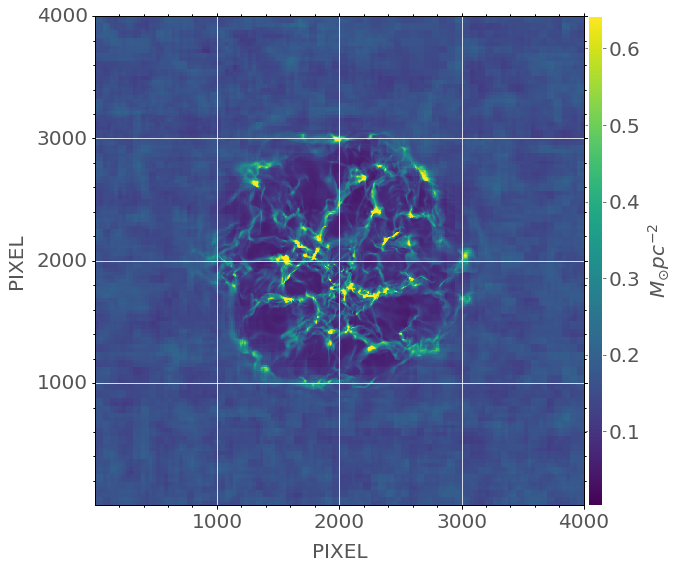}}
    \subfloat[]{\includegraphics[width=0.33\textwidth,height=0.23\textheight,trim={0cm 2cm 0cm 0cm},clip]{full_definition_from_SKIRT_t0125.png}}
    \subfloat[]{\includegraphics[width=0.33\textwidth,height=0.23\textheight,trim={0cm 2cm 0cm 0cm},clip]{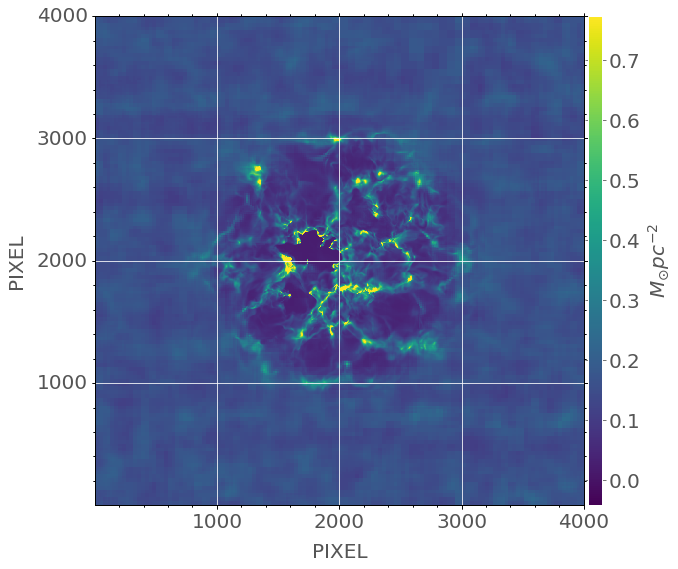}}\\
    \caption{
    {\it Top row}: Raw SEDs for the radiative transfer models when we use only stars (red line) or add the ISRF (blue line) as sources of radiation. The black dashed line represents the input SED of the ISRF. 
    {\it Bottom row}: Dust grid column-density projection of the face-on line of sight. Columns are the results of snapshots at 0.8, 1.6, 3.3 Myr.}
    \label{fig:24_microns_bands_and_SEDS}
\end{figure*}
}
%check 
\subsection{Comparison to Galactic star-forming regions}
Although the hydrodynamical simulation analyzed in this work does not attempt to reproduce any specific real region, given the physics it includes and characteristics such as surface density and filamentary clump morphology, we expect it to share similar characteristics to those of massive star-forming regions observed in the Galaxy \citep[e.g.,][]{LiuBaobab2012,galvan2013muscle,Motte2022,Traficante2023}. 

The first snapshot does not reproduce the typical SED of  young, massive star formation regions such as those in \citet{lin2017cloud}, even after the inclusion of the ISRF. Later snapshots do enter the range where they are comparable to IRDCs and the Orion star formation region 
\citep[e.g.,][]{lada2010star,Stutz18,Roman2019}. Figure  \ref{fig:binder_skirt_yuxin} shows this comparison with the Orion nebula SED provided in \citet{binder2018multiwavelength}. However, an important caveat is that the Orion SED was measured in a radius of a few pc around the Trapezium nebula, whereas the aperture of the observed synthetic SED has a size of 100 pc. Therefore the real Orion star formation region appears to be much more concentrated than the simulated cloud. 
Also, in general the SEDs of our simulation peak toward longer wavelengths as compared to the sample of galactic star-forming regions.

Therefore, more realistic hydrodynamical simulations should have a similar star formation rate but over a smaller volume. This will also allow to heat dust to higher temperatures, and produce SEDs that peak at about $100~\mu$m, as observed.

As mentioned in Section \ref{subsec:sinks}, the sink particles are populated in a stochastically-sampled way using the IMF as a probability density function, and from the different realizations we used the one that gave us the median bolometric luminosity. Since our cloud appears to be colder than the Galactic sample, we performed a radiative transfer model for the snapshot at 3.3 Myr using the realization having the maximum bolometric luminosity. 
This extreme model has a relative excess of massive stars, and a total input stellar luminosity of $10^7 \ L_\odot$ (see Appendix \ref{appendix_massive_stars}), which is three orders of magnitude higher than the model with median luminosity (see Table \ref{tab:temp_weighed_mass}). This extra energy produces an SED closer to the one of the W49A cloud, one of the most luminous star-formation regions in the Galaxy with $L_\mathrm{FIR} > 10^7~L_\odot$ \citep{LinYuxin2016}.  
The above result illustrates the dramatic effect that an excess of massive stars, either due to stochastical sampling \citep{OrozcoDuarte2022} or an intrinsically top-heavy IMF \citep{Schneider2018,Pouteau2022} can have in the observational appearance of molecular clouds.

%%%%%%%%%%%%%%%%%%%%%%%%%%%%
\subsection{Testing MBB fitting}

In Section \ref{sec:results}
we modeled the dust emission as a modified black body with a fixed emissivity index $\beta$, using 
both integrated and spatially resolved fluxes.

First, we used a single temperature MBB to reproduce the spatially-integrated \textit{Herschel} observations in the 160 to 500 $\mu$m range. This model  yielded a temperature averaged over the last four snapshots of 17 K, and a total dust mass of $\sim 430~M_\odot$ \ and $\sim 490~ M_\odot$ for the face-on and edge-on views, respectively. 
Therefore, the model underestimated the intrinsic dust mass by about a factor $\times 2$.
We also tested the MBB fitting by adding the $100~ \mu$m photometric point.
This fit yielded a dust temperature of  $\sim 20$ K, which resulted in a decrease of the dust mass by $\sim 100~ M_\odot$ with respect to the previous fit, for both lines of sight.
The MBB fits for this case tend to slightly underestimate the emission at the longer wavelengths. 
This suggests that assuming dust at a single temperature is not an optimal assumption, as this model would not properly represent variations in the line of sight. 

Since dust properties are luminosity weighted in the line of sight, and the IR emission tends to be dominated by the warmer dust \citep[see Section \ref{sec:intrinsic_mass}, and, e.g., ][]{Malinen_2011,Ysard2012A&A}, in this work we also use a 2-temperature MBB model using the six bands of \textit{Herschel} in order to characterize both the warm and cold dust components. We found that, although the warm dust component (with temperatures of $\sim 30$ K)  accounts for a very small fraction ($\leq 5~\%$) of the total mass recovered by the fit, the addition of a second component allows the colder component to better recover the emission at longer wavelengths. 
Therefore, the total dust using a 2-temperature MBB is closer to the intrinsic mass of the simulation. 
We suggest that whenever enough data points from the mid- to the far-IR are available, the best option is use two-component models  \citep[e.g.,][]{Immer2014}. While this is still a  simplified approach, it seems to be the best compromise between the number of data points and number of parameters.

We also explored the consequences of performing the MBB fitting in a spatially-resolved way. This has the advantage of partially alleviating the effect of line of sight variations, at least compared to spatially-integrated fluxes.
Furthermore, with this type of analysis we can follow the evolution of both the temperature and the surface density in our models.
From a comparison of the integrated and spatially-resolved results (see Table \ref{tab:fits_results}), we conclude that the latter, using a 2-temperature MBB, is the 
one that recovers a total mass closer ($\sim66 ~\%$)
to the intrinsic dust mass in the simulation grid. 

As we mentioned in Section \ref{subsec:discution_isrf}, \texttt{SKIRT} calculates the dust temperature self-consistently, and we can not define a base  temperature for the dust prior to the radiative transfer simulation.
In an analysis similar to ours,  \cite{koepferl17a} found that in their model without background, in which they couple the dust temperature to an isothermal field at  $T_{\mathrm{iso}} = 18$ K, the total mass recovered in their pixel-by-pixel analysis is  very close to the intrinsic mass in their simulations. 
The median ratio and median absolute deviation of their recovered to intrinsic column densities is $1.12\pm0.28$. 
In our case, for the face-on line of sight, we find a ratio and deviation of $0.58 \pm 0.12$ for the resolved, single-component MBB fitting, and $0.66 \pm 0.13$ for the case of two-component MBBs, also in a resolved way. Both values are averaged over the last four snapshots. 

We further explored the fact that in the spatially resolved analysis the calculated total mass is closer to the intrinsic mass, but still underestimated.
We performed a test in which we reduce the size of the analysis box $M_{100pc}$ in the x-axis from 100 to 30 pc, or about the size of densest part of the cloud in this axis (see Figure \ref{fig:mosaico_edge_total_mass}). 
The total mass inside this new box, averaged in the last four snapshots, is $563~M_\odot$. 
For the case of this reduced box, the ratio of the dust mass recovered by resolved 2-temperature MBB fitting is $\sim 1.04$ times the intrinsic one. This proves that the origin of the ``missing mass'' is that our analysis does not recover the coldest dust in the simulation grid.

\subsection{SFR}
Individual star forming regions are in general characterized by different stellar contents and SFR values, this being due to their different initial conditions and evolutionary stage 
\citep[e.g.,][]{lada2010star,binder2018multiwavelength,Motte2022}. 

We now check how the SFR varies among different parts of our simulated cloud, and how this broadly compares with values observed in actual massive star formation regions. The SFR value calculated from different tracers, in our case, monochromatic and total IR luminosities, 
will generally depend on the evolutionary stage of the cloud, and on where this luminosity is measured. We calculated SFRs using the calibration coefficients $a$ found in Section \ref{subsec:SFR_calibrations}, using apertures of different sizes centered in two sub-regions. These sub-regions represent contrasting scenarios in the 3.3 Myr snapshot. The first one, region A, is centered on the massive star-forming region which has an \textsc{Hii} region in expansion (R1 in Fig. \ref{fig:mosaico_face_total_mass}). The second, region B, is centered on a young star-forming region that appears as an over-density at 5.7 Myr (magenta ellipse in  Figure \ref{fig:mosaico_face_total_mass}). 

\begin{figure*}
    \centering
    \includegraphics[height=0.3\textheight, width =  \textwidth]{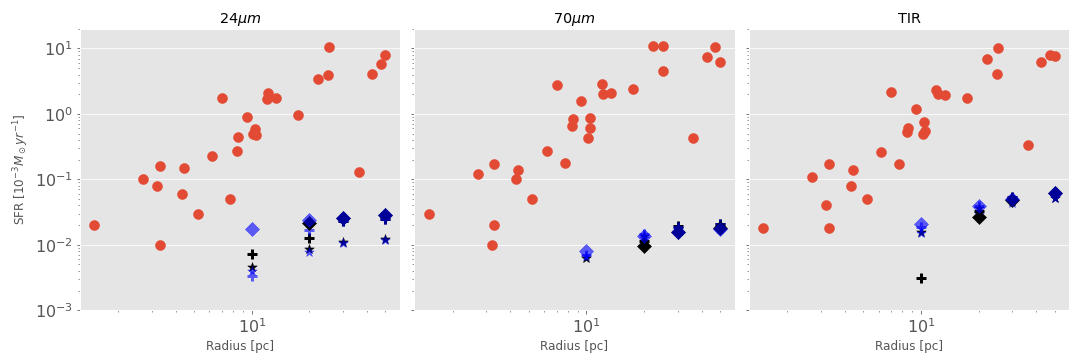}
    \caption{SFR plotted versus aperture size of the observation.
    The red dots represent SFR values taken from the \cite{binder2018multiwavelength} sample.
    The other symbols represent the inferred SFRs for simulation snapshots: $\star = $ 3.3 Myr, $+ =$ 5.7 Myr, and $\Diamond = $ 7.6 Myr.
    Black symbols represent the synthetic observations of region A (\textsc{Hii} region). Blue symbols represent region B (younger, triggered star formation).}
    \label{fig:SFR_vs_aperture}
\end{figure*}

In Figure \ref{fig:SFR_vs_aperture} we plot the SFRs for a Galactic sample, whose data have been taken with apertures of different sizes \citep{binder2018multiwavelength}, along with the SFRs for our synthetic observations. We use apertures of 10, 20, 30, and 50 pc, at 3.3 Myr, 5.7 Myr, and 7.6 Myr. We consider the monochromatic 24 and 70 $\mu$m, and the total IR luminosity. The constants $a$ that we use are those averaged over 100 Myr. It is seen that, for a given aperture radius, the inferred SFRs for the simulation are systematically smaller than in observations. This again suggests that star formation in the simulated cloud is more scattered than in real clouds.

Unsurprisingly, SFRs calculated at all radii for region A are generally smaller than those calculated for region B. This is because region A is progressively getting depleted from dust due to ionization feedback. In contrast, in region B the amount of dust is steadily increasing due to the concurrent action of the nearby \textsc{Hii} regions, which trigger star formation at this location. 

The 24 $\mu$m tracer is the one that suffers larger changes due to both the evolutionary stage of the region and, to some degree, the aperture size. This is likely due to the fact that at this wavelength the emission is dominated by the warmer dust which is located closer to the young stars, and that this brighter, warmer dust occupies a smaller volume compared to the colder dust. Therefore its spatial configuration and resulting flux evolves faster in time. As for the 70 $\mu$m and total IR tracers, there is only a weak dependence of the \ SFR as a function of the aperture used for the flux measurement. Furthermore, the value calculated using these two tracers shows almost no differences as a function of time, with the only exception of the one calculated in the smaller region of the 5.7 Myr snapshot. Note also that for the smallest  aperture centered in region A, the inferred SFR is very small, thus is not shown in Figure \ref{fig:SFR_vs_aperture}. 

\section{Conclusions} \label{sec:conclusions}

In this work we have presented and analyzed post-processing Monte Carlo dust radiative transfer, made with \texttt{SKIRT}, of the hydrodynamical simulation with ionizing feedback presented by \citet{ZA+19}, which was created in the context of the global hierarchical collapse scenario \citep{vazquez2019global}. The synthetic observations were analyzed using common observational techniques, including photometry and morphological analysis, as well as integrated and resolved SED fitting. 

A crucial ingredient in the creation of the synthetic observations is the choice of the primary sources of photons. To consider the effects of stochasticity in the stellar populations that are represented by sink particles, we sampled each of them with many realizations of an IMF and selected those with the median luminosity. The effects of a different selection were explored and can be dramatic. 
Another key aspect was the inclusion of a prescription for the interstellar radiation field, which allowed to pre-heat the dust over the entire cloud to realistic levels prior to significant star formation. 

The infrared appearance of the simulated cloud evolves from a quiescent region, analogous to IRDCs, to active star formation with the presence of bubbles and ridges caused by the expansion of \textsc{Hii} regions. This evolution occurs over a total interval of about 8 Myr after the onset of star formation, but the global IR SED settles rather quickly after 3 Myr. The final luminosity of the cloud is comparable to that of the Orion nebula cluster, but spread over a much larger area. 

We used the synthetic observations to test the effectiveness of standard techniques to recover properties such as dust mass and temperature. We performed modified black body fitting with one and two temperature components, both in the spatially-integrated SEDs and in a resolved fashion. MBB fitting systematically underestimates the intrinsic dust mass in the simulation grid by about a factor $\times 2$. Pixel-by-pixel fitting recovers about $70\%$ of the dust mass. The ``missing mass'' consists of the coldest dust with temperatures significantly below 10 to 15 K, whose thermal emission does not contribute to the flux of the cloud, even at far-IR wavelengths. We believe that this effect is real and could affect observations, insofar clouds have some very cold dust. 

Finally, we tested observational calibrations of the global star formation rate of the cloud based on monochromatic fluxes at 24 and 70 $\mu$m, as well as total IR emission, obtaining results consistent with observations of a Milky Way sample reported by \citet{binder2018multiwavelength}. However, the star formation rate in the simulated cloud -- as mentioned before for the case of the luminosity -- appears to be less concentrated than in observed star forming clouds. 
This suggests that the prescription for the assembly mechanism responsible for cloud formation and evolution in the simulation is not capable of bringing enough gas into a small enough region. 
Future simulations within the GHC scenario will explore this issue, which likely requires further gravitational focusing from even larger scales.

Regarding the comparison of simulations to observations, a step forward from the work presented in this paper will be to analyze a sample of simulated clouds spanning the relevant range of formation mechanisms and physical conditions known to exist in molecular clouds. Ideally, the simulated clouds should form self-consistently from larger (galactic) scales \citep[e.g.,][]{Walch2015,smith2020cloud}, and include realistic prescriptions for their turbulence and magnetic fields. Also, the relevant sources of stellar feedback (\textsc{Hii} regions, radiation pressure onto dust, stellar winds, bipolar outflows) should ideally be included. Recent efforts toward that direction have been presented by \citet{Grudic2022}. This is a challenging issue, and typically simulations are only able to include some but not all of these processes \citep[e.g.,][]{Peters2011,Krumholz2012,Dale2014}. 
We plan to present further work on the comparison of synthetic and real observations in the future.

\smallskip

\acknowledgments
JJD and RGM acknowledge support from UNAM-PAPIIT project IN108822, and from CONACyT Ciencia de Frontera project ID: 86372. RGM also acknowledges support from the AAS Chr\'etien International Research Grant. JJD and JF acknowledge support from UNAM-PAPIIT project IN111620. The authors thank the anonymous referee for a timely and useful report. 

\smallskip

\software{\texttt{astropy} \citep{astropy:2022},  
          \texttt{SKIRT} \citep{camps2015skirt,camps2020SKIRT}, 
          \texttt{LMFIT} \citep{matt_newville_2021_5570790},
          \texttt{IMF} (\url{https://github.com/keflavich/imf}
          }

\appendix

\section{Example Flat SED Fitting}
\label{sec:app_A} 
\begin{figure}[!h]
    \centering
    \includegraphics[width = 0.4\textwidth, height = 0.3\textheight]{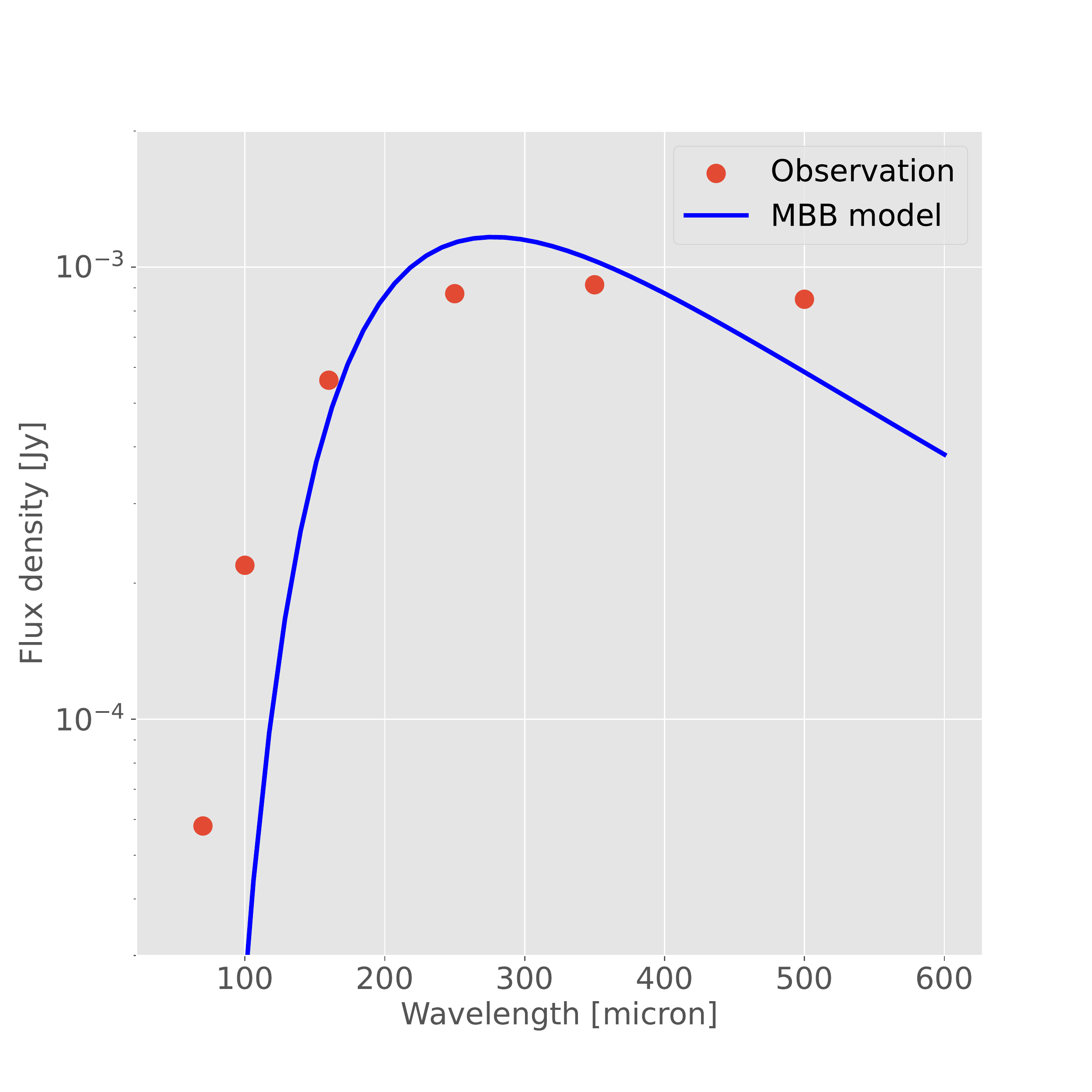}
    \caption{Representative SED of a pixel with a poor fit in the first two timesteps of Figure \ref{fig:mosaico_face_total_mass}. The blue line represents the MBB emission with the temperature and mass that best fit the observed fluxes (red dots). The flat SED observed at the two earlier snapshots was already discussed for the integrated SEDs in section \ref{subsec:Observed_SED_comparision_observations}, and is also seen at the pixel scale. }
    \label{fig:bad_pixels_sample}
\end{figure}

As pointed out in Section \ref{pix_by_pix}, in the two youngest snapshots of the simulation, dust emission in some pixels is not adequately modelled by a single modified black body. Figure \ref{fig:bad_pixels_sample} shows a representative example.    

\section{Synthetic Evolution in Edge-On View}
\label{sec:app_B} 
\begin{figure*}[!h]
    \centering
    \includegraphics[width=\textwidth,height=0.5\textheight,trim={2cm 2cm 0cm 0cm},clip]{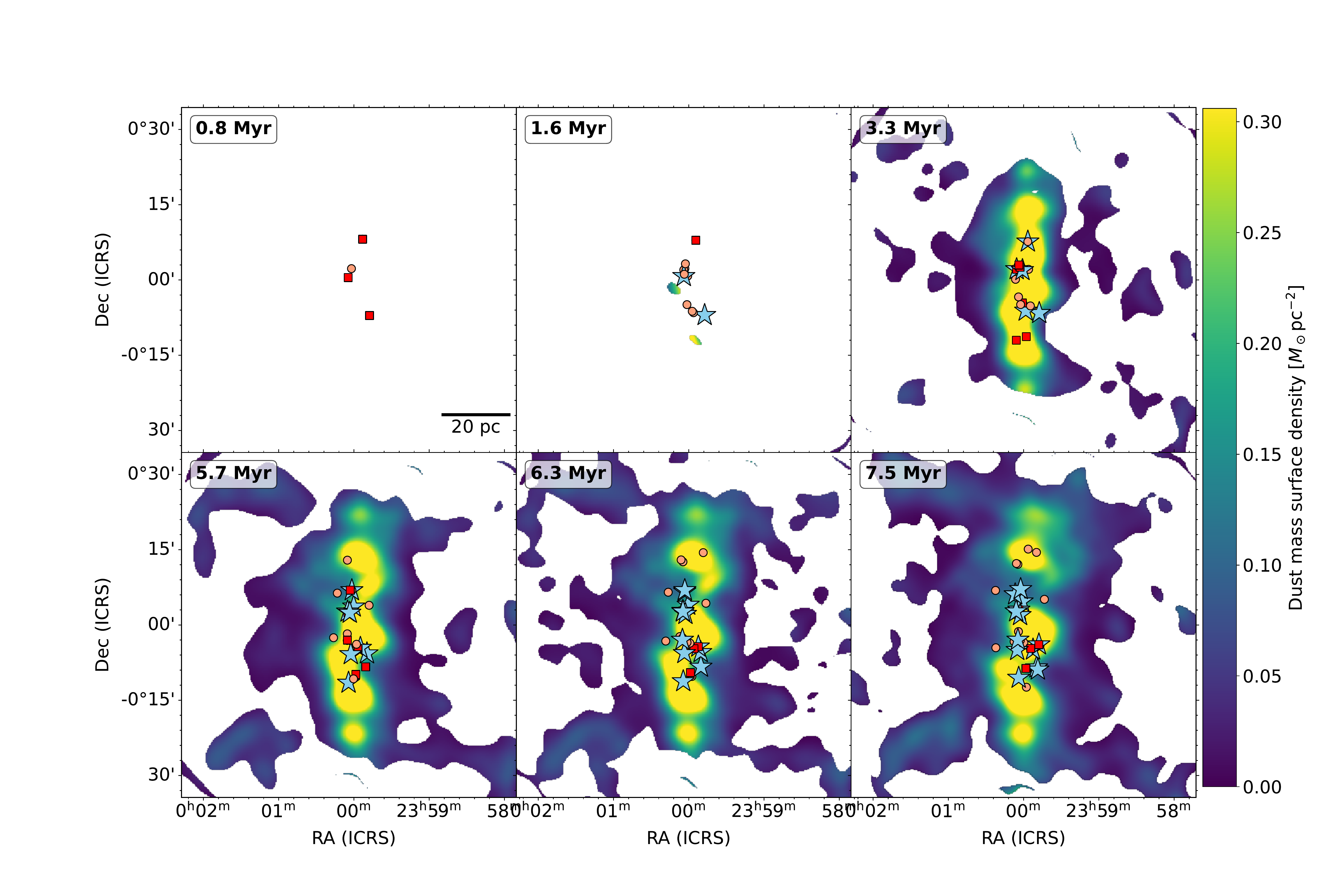}
    \caption{
    Dust surface density maps  ($M_\odot ~ \mathrm{pc^{-2}}$) resulting from the spatially resolved fitting to the edge-on synthetic observations at various timesteps. The markers represent the projected sink positions, in blue stars for the sinks with masses $\geq 100~ M_\odot$, in red squares for sinks with masses between $ 50~ M_\odot$ and $100~ M_\odot$, and in orange circles the sinks with  $ \leq 50~ M_\odot$.}
    \label{fig:mosaico_edge_total_mass}
\end{figure*}

\begin{figure*}[!h]
    \centering
    \includegraphics[width=\textwidth,height=0.5\textheight,trim={2cm 2cm 0cm 0cm},clip]{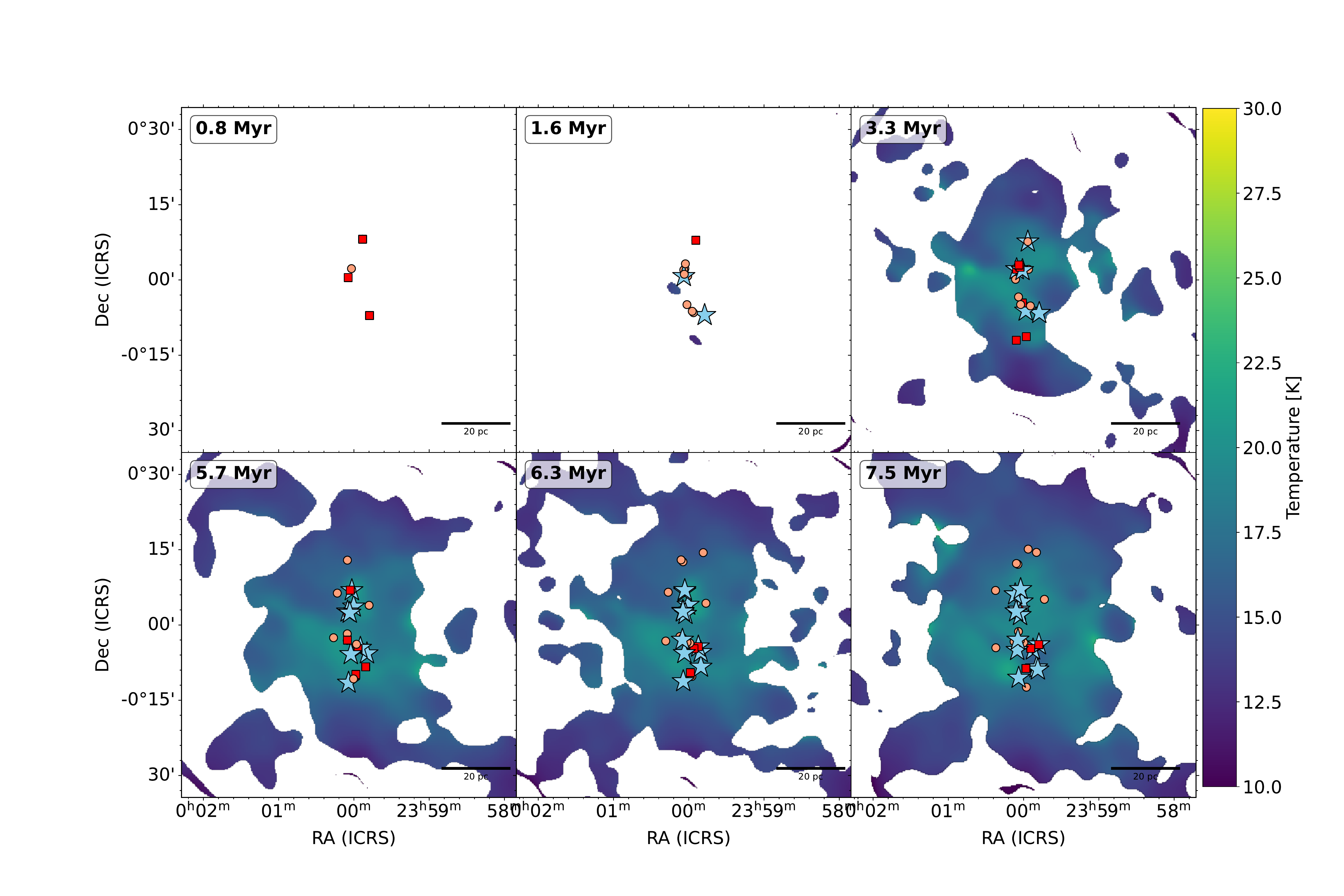}
    \caption{Dust temperature maps in Kelvin resulting from the spatially resolved fitting to the edge-on synthetic observations at various timesteps. 
    Labels and symbols are the same as in Figure \ref{fig:mosaico_edge_total_mass}.  }
    \label{fig:mosaico_edge_total_temperature}
\end{figure*}

Here we show the evolution of the recovered dust surface density (Figure \ref{fig:mosaico_edge_total_mass}) and temperature (\ref{fig:mosaico_edge_total_temperature}) when the simulated cloud is viewed from an edge-on orientation.

\section{Exploring the effect of an excess of massive stars}\label{appendix_massive_stars}

In the building of the input SEDs for each sink, we have explored about 200 different ways to distribute the total sink mass into individual stars, using a stochastic sampling of the Kroupa IMF. Then, we have calculated the bolometric luminosity for each of these realizations, and we have used as input spectra for the sinks the ones with median luminosity. 
In this appendix, we test the effects of taking the spectra with the maximum luminosity, which contain more massive stars with respect to the ones with the median luminosity. An excess of massive stars could be obtained due to real stochasticity in the forming cluster, or by an intrinsically top-heavy stellar IMF. 

\begin{figure}[!h]
    \centering
    \includegraphics[width=0.5\textwidth,height=0.35\textheight]{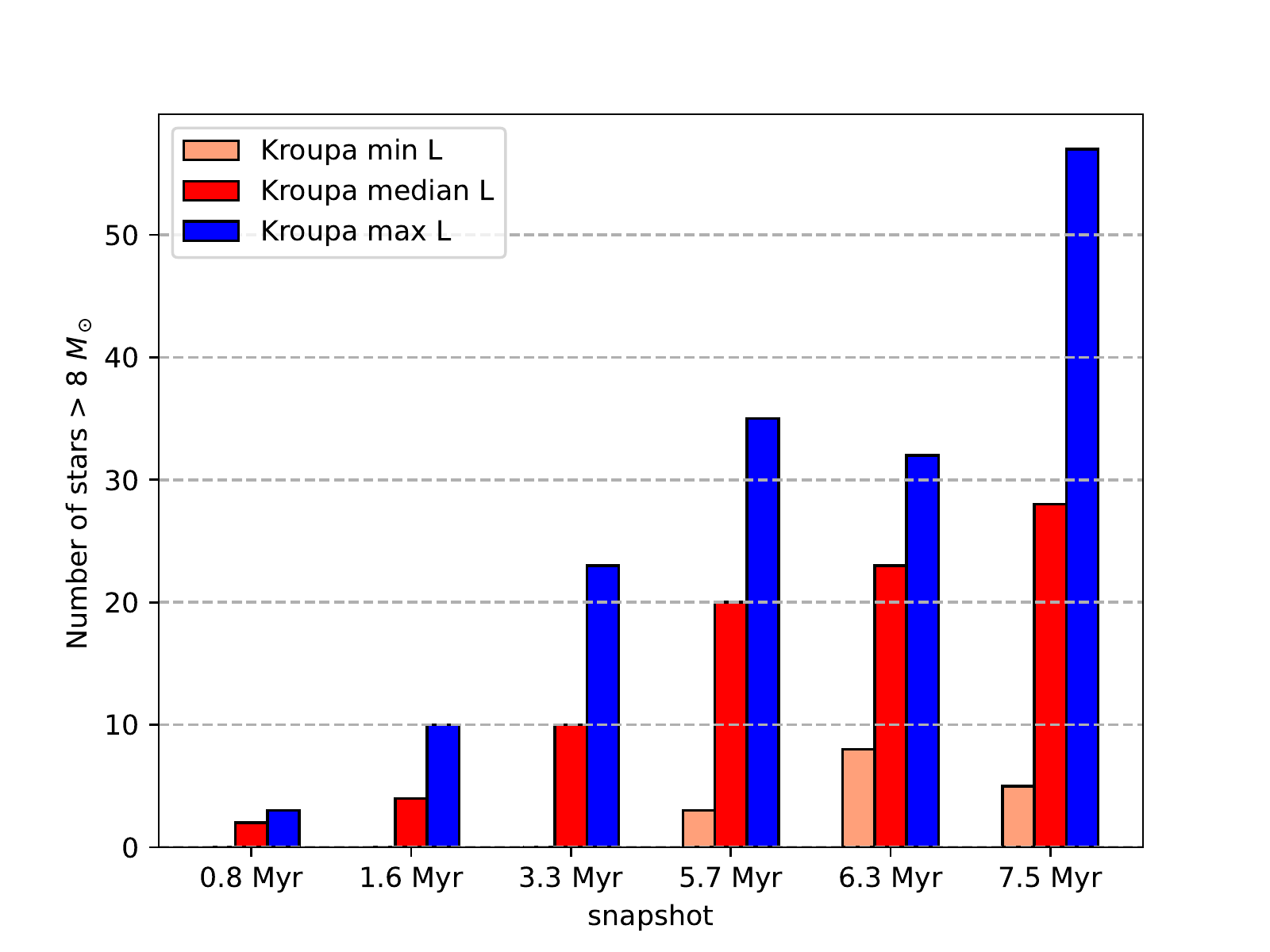}
    \caption{Histogram of the total number of stars with masses higher than $8~M_\odot$, for the minimum, median, and maximum luminosity realizations resulting from the IMF stochastic sampling.}
    \label{fig:stars_distributions_min_vs_median_vs max}
\end{figure}

\begin{figure}[!h]
    \centering
    \includegraphics[width=0.7\textwidth,height=0.35\textheight]{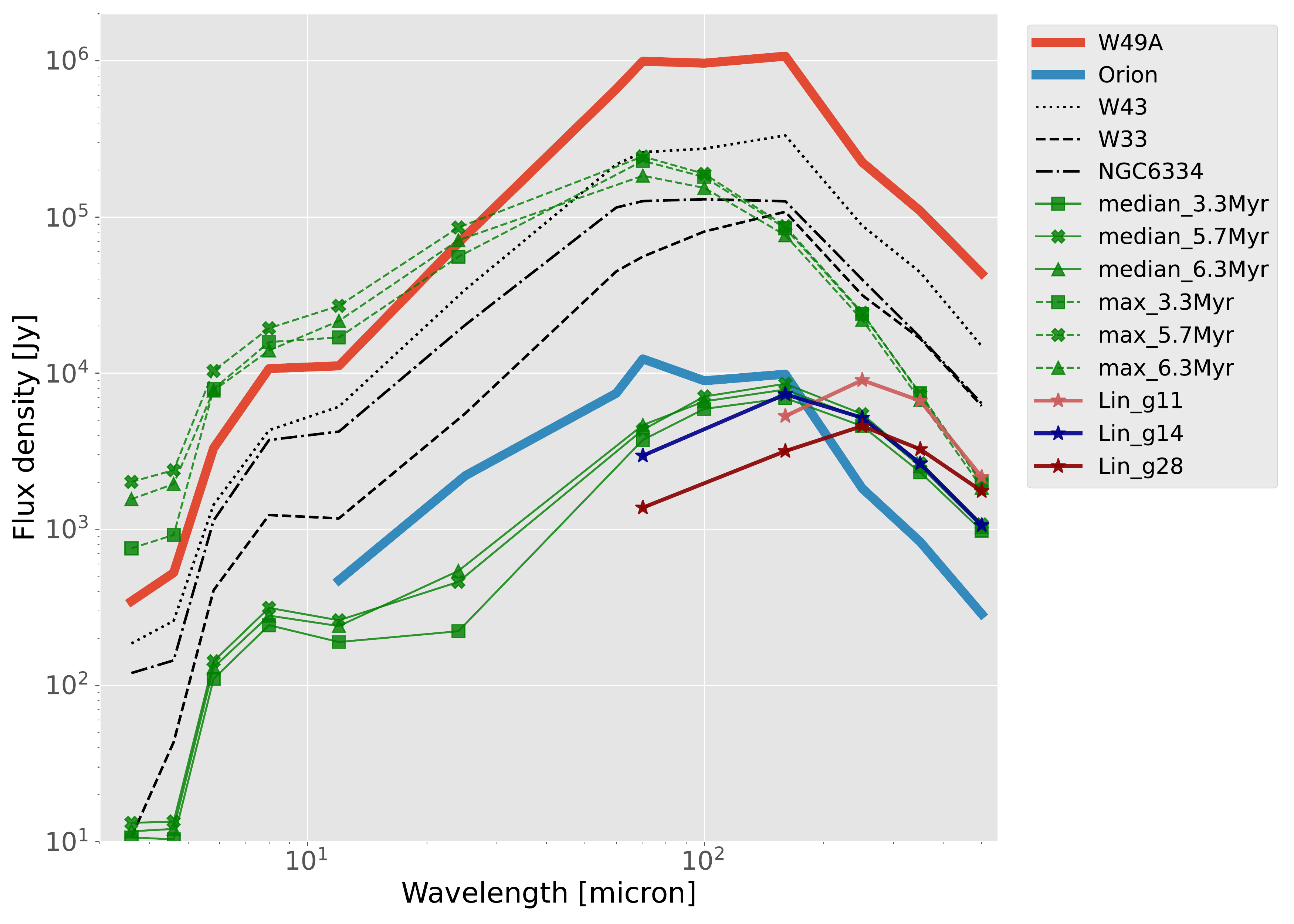}
    \caption{Similar to Figure \ref{fig:binder_skirt_yuxin}, but the median-luminosity models are only plotted for the snapshots at 3.3, 5.7, and 6.3 Myr (face-on, solid green lines). The corresponding maximum-luminosity models are plotted with green dashed lines.}
    \label{fig:max_vs_median_seds_comparision}
\end{figure}

 Figure \ref{fig:stars_distributions_min_vs_median_vs max} shows the number of massive ($M > 8~M_\odot$) stars for each snapshot and for three different realizations of the stochastic IMF sampling: minimum, median, and maximum luminosity. 
In Figure \ref{fig:max_vs_median_seds_comparision} we compare the SEDs of the face-on, median-luminosity models (also shown in Figure \ref{fig:binder_skirt_yuxin}) with those of maximum luminosity. We performed this test for the snapshots at 3.3, 5.7, and 6.3 Myr. One of the most notable differences, in addition to the much larger luminosities, is that the peak of the SED is displaced to shorter ($\sim 70~ \mu$m) wavelengths.

\bibliography{references}{}
\bibliographystyle{aasjournal}

\end{document}